\PassOptionsToPackage{unicode}{hyperref}
\PassOptionsToPackage{hyphens}{url}
\PassOptionsToPackage{dvipsnames,svgnames,x11names}{xcolor}
\documentclass[
]{article}
\usepackage{amsmath,amssymb}
\usepackage{lmodern}
\usepackage{iftex}
\ifPDFTeX
  \usepackage[T1]{fontenc}
  \usepackage[utf8]{inputenc}
  \usepackage{textcomp} 
\else 
  \usepackage{unicode-math}
  \defaultfontfeatures{Scale=MatchLowercase}
  \defaultfontfeatures[\rmfamily]{Ligatures=TeX,Scale=1}
\fi
\IfFileExists{upquote.sty}{\usepackage{upquote}}{}
\IfFileExists{microtype.sty}{
  \usepackage[]{microtype}
  \UseMicrotypeSet[protrusion]{basicmath} 
}{}
\makeatletter
\@ifundefined{KOMAClassName}{
  \IfFileExists{parskip.sty}{%
    \usepackage{parskip}
  }{
    \setlength{\parindent}{0pt}
    \setlength{\parskip}{6pt plus 2pt minus 1pt}}
}{
  \KOMAoptions{parskip=half}}
\makeatother
\usepackage{xcolor}
\usepackage[margin=1in]{geometry}
\usepackage{color}
\usepackage{fancyvrb}

\DefineVerbatimEnvironment{Highlighting}{Verbatim}{commandchars=\\\{\}}
\usepackage{framed}
\definecolor{shadecolor}{RGB}{248,248,248}
\newenvironment{Shaded}{\begin{snugshade}}{\end{snugshade}}

\newcommand{\AttributeTok}[1]{\textcolor[rgb]{0.77,0.63,0.00}{#1}}

\newcommand{\CommentTok}[1]{\textcolor[rgb]{0.56,0.35,0.01}{\textit{#1}}}

\newcommand{\DecValTok}[1]{\textcolor[rgb]{0.00,0.00,0.81}{#1}}

\newcommand{\FloatTok}[1]{\textcolor[rgb]{0.00,0.00,0.81}{#1}}
\newcommand{\FunctionTok}[1]{\textcolor[rgb]{0.00,0.00,0.00}{#1}}

\newcommand{\NormalTok}[1]{#1}

\newcommand{\SpecialCharTok}[1]{\textcolor[rgb]{0.00,0.00,0.00}{#1}}

\newcommand{\StringTok}[1]{\textcolor[rgb]{0.31,0.60,0.02}{#1}}

\usepackage{longtable,booktabs,array}
\usepackage{calc} 
\usepackage{etoolbox}
\makeatletter
\patchcmd\longtable{\par}{\if@noskipsec\mbox{}\fi\par}{}{}
\makeatother
\IfFileExists{footnotehyper.sty}{\usepackage{footnotehyper}}{\usepackage{footnote}}
\makesavenoteenv{longtable}
\usepackage{graphicx}
\makeatletter
\def\maxwidth{\ifdim\Gin@nat@width>\linewidth\linewidth\else\Gin@nat@width\fi}
\def\maxheight{\ifdim\Gin@nat@height>\textheight\textheight\else\Gin@nat@height\fi}
\makeatother
\setkeys{Gin}{width=\maxwidth,height=\maxheight,keepaspectratio}
\makeatletter
\def\fps@figure{htbp}
\makeatother
\providecommand{\tightlist}{%
  \setlength{\itemsep}{0pt}\setlength{\parskip}{0pt}}
\newlength{\cslhangindent}
\newlength{\csllabelwidth}
\newlength{\cslentryspacingunit} 
\newenvironment{CSLReferences}[2] 
 {
  \setlength{\parindent}{0pt}
  \ifodd #1
  \let\oldpar\par
  \def\par{\hangindent=\cslhangindent\oldpar}
  \fi
  \setlength{\parskip}{#2\cslentryspacingunit}
 }%
 {}
\usepackage{calc}

\usepackage{pdflscape}
\newcommand{\blandscape}{\begin{landscape}}
\newcommand{\elandscape}{\end{landscape}}
\ifLuaTeX
  \usepackage{selnolig}  
\fi
\IfFileExists{bookmark.sty}{\usepackage{bookmark}}{\usepackage{hyperref}}
\IfFileExists{xurl.sty}{\usepackage{xurl}}{} 
\hypersetup{
  pdftitle={Understanding Complex Patterns in Social, Geographic, and Economic Inequities in COVID-19 Mortality at the County Level in the US Using Generalized Additive Models},
  pdfauthor={Christian Testa1,},
  colorlinks=true,
  linkcolor={black},
  filecolor={Maroon},
  citecolor={Blue},
  urlcolor={blue},
  pdfcreator={LaTeX via pandoc}}

\title{Understanding Complex Patterns in Social, Geographic, and Economic Inequities in COVID-19 Mortality at the County Level in the US Using Generalized Additive Models}
\author{Christian Testa\textsuperscript{1,*}}
\date{November 29, 2022}

\begin{document}
\maketitle
\begin{abstract}
I present three types of applications of generalized additive models (GAMs) to
COVID-19 mortality rates in the US for the purpose of advancing methods to
document inequities with respect to which communities suffered disproportionate
COVID-19 mortality rates at specific times during the first three years of the
pandemic. First, GAMs can be used to describe the changing
relationship between COVID-19 mortality and county-level covariates
(sociodemographic, economic, and political metrics) over time. Second, GAMs can
be used to perform spatiotemporal smoothing that pools information over time and
space to address statistical instability due to small
population counts or stochasticity resulting in a smooth, dynamic latent
risk surface summarizing the mortality risk associated with geographic locations over
time. Third, estimation of COVID-19 mortality associations with county-level covariates
conditional on a smooth spatiotemporal risk surface allows for more rigorous consideration
of how socio-environmental contexts and policies may have impacted COVID-19 mortality.
Each of these approaches
provides a valuable perspective to documenting inequities in
COVID-19 mortality by addressing the question of which populations have suffered
the worst burden of COVID-19 mortality taking into account the nonlinear
spatial, temporal, and social patterning of disease.
\newline

\noindent Abbreviations used: United States (US), Coronavirus Disease 2019 (COVID-19), Generalized
Additive Model (GAM), Centers for Disease Control and Prevention (CDC), Index of Concentration at the Extremes (ICE), Confidence Interval (CI)
\end{abstract}

\textsuperscript{1} Department of Social and Behavioral Sciences, Harvard T.H. Chan School of Public Health

\textsuperscript{*} Correspondence: \href{mailto:ctesta@hsph.harvard.edu}{Christian Testa \textless{}\href{mailto:ctesta@hsph.harvard.edu}{\nolinkurl{ctesta@hsph.harvard.edu}}\textgreater{}}

\hypertarget{introduction}{%
\section{Introduction}\label{introduction}}

As we enter the third winter with the novel coronavirus disease
COVID-19 in the United States, evidence documenting the intense disparities in
COVID-19 mortality rates comparing socially advantaged and disadvantaged
populations continues to mount. Eliminating inequities in health outcomes has
been stated as a major policy goal of the Biden administration
(\protect\hyperlink{ref-the_white_house_executive_2021}{The White House, 2021}, \protect\hyperlink{ref-the_white_house_advancing_2022}{2022}), representing
a revitalized commitment to health equity and underlining the importance of adequate
data reporting systems that report timely estimates of prevalent health inequities. To this
end, I use generalized additive models (GAMs) as a flexible regression
framework to illustrate the evolving roles and relationships sociodemographic,
geographic, and economic conditions have with respect to trends in COVID-19
mortality. The code, data, and documentation necessary to reproduce the analyses
contained in this paper are online and free to access at \url{https://github.com/ctesta01/covid.gradient.estimation}.

\hypertarget{background}{%
\subsubsection{Background}\label{background}}

Having passed over 1 million COVID-19 deaths in the United States in 2022
(\protect\hyperlink{ref-donovan_us_2022}{Donovan, 2022}), and facing uncertain prospects for the third COVID-19 winter
looming even as new iterations on the COVID-19 vaccines become available, it
remains critical that inequities in COVID-19 outcomes are documented and
analyzed to reckon with the unjust and unfair burden of preventable illness.
Even though the first vaccines were granted emergency use authorization by the
U.S. Food and Drug Administration in 2020 (\protect\hyperlink{ref-mayo_clinic_history_2022}{Mayo Clinic, 2022}), with the
first shots going in arms in December 2020, COVID-19 is still continuing to
cause hundreds of deaths a day in the US in the fall of 2022
(\protect\hyperlink{ref-noauthor_united_2022}{{``United {States} {COVID} - {Coronavirus} {Statistics} - {Worldometer},''} 2022}). The new bivalent vaccines released at the end of
August 2022 contain mRNA sequences from both the original strain as well as the
recently emergent BA.4 and BA.5 lineages in an effort to make the nation's
immunity more up-to-date and robust against the myriad of phylogenetic
directions the COVID-19 virus is evolving to explore
(\protect\hyperlink{ref-office_of_the_commissioner_coronavirus_2022}{Office of the Commissioner, 2022}). Despite the updated bivalent
boosters representing a significant step forward in prevention strategy, less
than 4\% of eligible Americans had taken the booster in the first month after it
became available (\protect\hyperlink{ref-bendix_less_2022}{Bendix, 2022}; \protect\hyperlink{ref-lambert_most_2022}{Lambert, 2022}). As such, and with an
enduring history of inequities in health care access in the US
(\protect\hyperlink{ref-bailey_how_2021}{Bailey et al., 2021}; \protect\hyperlink{ref-blendon_inequities_2002}{Blendon et al., 2002}; \protect\hyperlink{ref-carpenter_health_2021}{Carpenter, 2021}; \protect\hyperlink{ref-chrisler_ageism_2016}{Chrisler et al., 2016}; \protect\hyperlink{ref-feldman_health_2021}{Feldman et al., 2021}; \protect\hyperlink{ref-okonkwo_covid-19_2021}{Okonkwo et al., 2021}; \protect\hyperlink{ref-ortega_ending_2021}{Ortega \& Roby, 2021}; \protect\hyperlink{ref-rapp_statelevel_2022}{Rapp et al., 2022}; \protect\hyperlink{ref-whitehead_outness_2016}{Whitehead et al., 2016}), it is clear that without further intervention not all
communities will be equally able to benefit from the updated vaccines and
inequities in COVID-19 illness and mortality may persist despite the
technological innovations in vaccine technology.

\hypertarget{the-role-of-geography-in-covid-19}{%
\subsubsection{The Role of Geography in COVID-19}\label{the-role-of-geography-in-covid-19}}

\begin{figure*}

{\centering \includegraphics[width=6in]{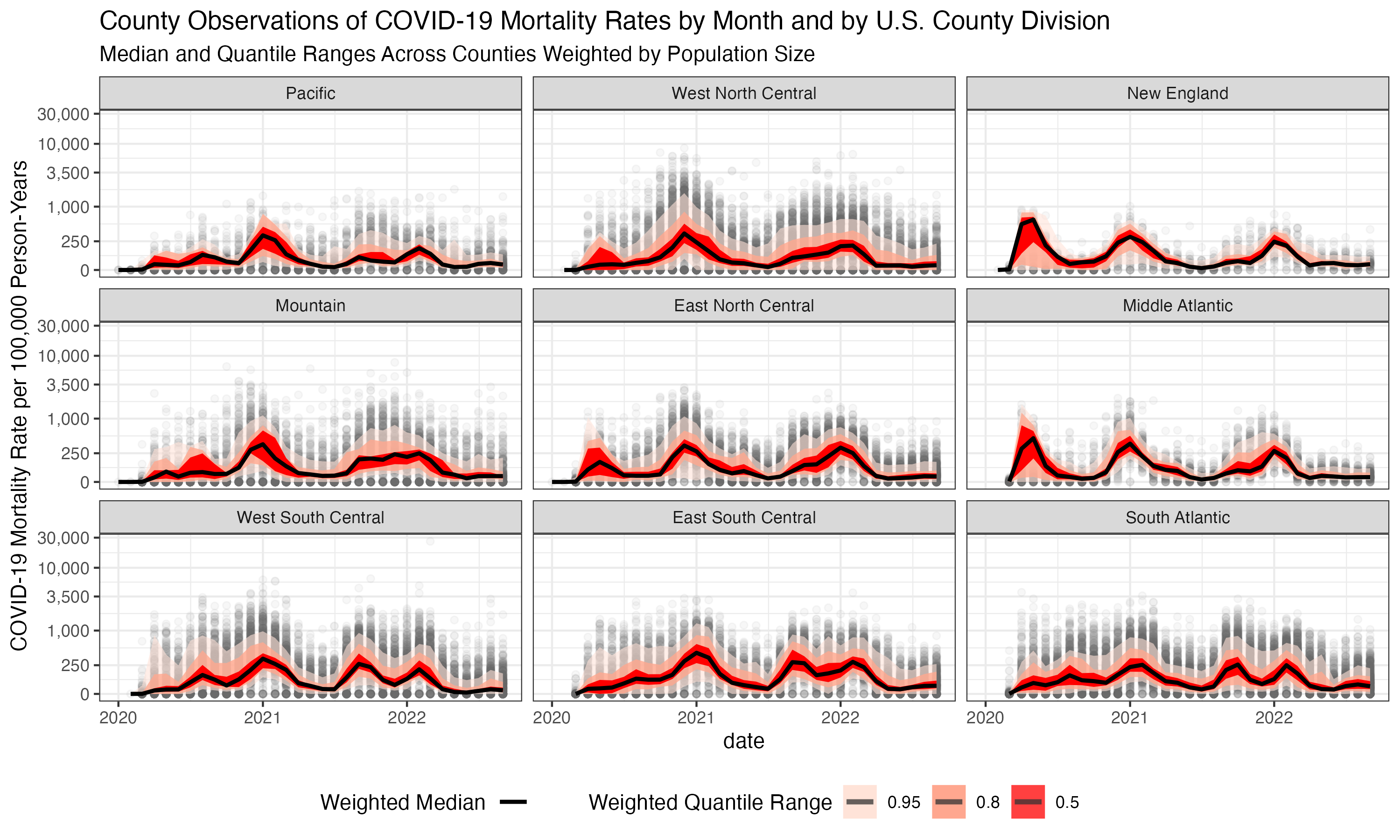}

}

\caption{Estimates of monthly COVID-19 mortality rates per 100,000 person-years by county organized by Census Division.  For each division, the median trendline and quantile ranges are shown weighted by county population size.}\label{fig:figure-county-trends}
\end{figure*}

Prior literature has demonstrated that the spread and impact of COVID-19 has
varied geography over time reflecting dynamics in the timing of introduction and
transmission events. Methods employed to highlight the geographic patterning in
COVID-19 outcomes have included quantile regression
(\protect\hyperlink{ref-sigler_socio-spatial_2021}{Sigler et al., 2021}), Besag-York-Molli\textquotesingle e mixed models
(\protect\hyperlink{ref-whittle_ecological_2020}{Whittle \& Diaz-Artiles, 2020}), spatial cluster analysis (\protect\hyperlink{ref-sugg_mapping_2021}{Sugg et al., 2021}),
geographically weighted regression (\protect\hyperlink{ref-mollalo_gis-based_2020}{Mollalo et al., 2020}; \protect\hyperlink{ref-park_covid-19_2021}{Park et al., 2021}), and others.

Figure \ref{fig:figure-county-trends} shows the monthly COVID-19 mortality
rates for the counties grouped within each of the nine U.S. Census Divisions
(\protect\hyperlink{ref-us_department_of_commerce_economics_and_statistics_administration_census_2000}{U.S. Department of Commerce Economics and Statistics Administration \& U.S. Census Bureau, 2000}).
The figure summarizes each division's median mortality rates
weighted by county population size.
Notably, the mortality associated
with the early surge of cases starting in New York City and spreading through New York,
New Jersey, and Massachusetts is visible in the Middle Atlantic and New England
division figures. The figure also illustrates how the first peak in the
mortality time-series for states in the Midwest (West North Central, East North
Central) occurred later, in late 2020 and going into early 2021.

In the US context, one of the key aspects to the geographic story of COVID-19's
spread and diffusion was the early surge of cases and epicenter in New York City
during March 2020 (\protect\hyperlink{ref-thompson_covid-19_2020}{Thompson, 2020}) followed by subsequent waves of cases in the
South and Midwest (\protect\hyperlink{ref-glenza_covid_2020}{Glenza, 2020}; \protect\hyperlink{ref-scott_these_2020}{Scott, 2020}; \protect\hyperlink{ref-shumaker_covid-19_2020}{Shumaker \& Wu, 2020}). As Park et al.~
stated summarized the trends in the US from March 2020 to May 2021, ``hot spots have shifted from densely populated cities and the states with a high percentage of socially vulnerable individuals to the states with relatively relaxed social distancing requirements, and then to the states with low vaccination rates'' (\protect\hyperlink{ref-park_covid-19_2021}{2021}).

When considering the drivers of the COVID-19 pandemic, it's necessary to note that
geography and social conditions are inextricably linked. In July 2021, the CDC
reported that ``the COVID-19 cumulative death rate in non-metropolitan areas has
exceeded that of metropolitan areas since December 2020,'' noting that of the
approximately 1/5th of Americans who live in rural areas, many ``are considered
highly vulnerable according to CDC's Social Vulnerability Index (SVI), which
includes factors such as housing, transportation, socioeconomic status, race,
and ethnicity'' (\protect\hyperlink{ref-cdc_location_2021}{CDC, 2021}). Moreover, rural communities often have
lower health insurance rates, higher disability rates, older populations, and
limited access to health care. One of CDC's Morbidity and Mortality Weekly
Reports found that vaccination against COVID-19 was lower in rural communities
than in urban communities between December 2020 and April 2021
(\protect\hyperlink{ref-murthy_disparities_2021}{Murthy, 2021}).

\hypertarget{the-social-determinants-of-covid-19-mortality}{%
\subsubsection{The Social Determinants of COVID-19 Mortality}\label{the-social-determinants-of-covid-19-mortality}}

Even since the beginning of the COVID-19 outbreak in the US, data reflected
sharp inequities in mortality rates. During January 22nd to May 5th 2020,
county COVID-19 mortality rates were 4.94 (95\% CI 4.78, 5.09) times higher in
counties in the highest quintile of percent People of Color (61\%-100\%) compared
to counties in the lowest quintile of percent People of Color (0\%-17.2\%)
(\protect\hyperlink{ref-chen_revealing_2021}{Chen \& Krieger, 2021}). This was not wholly unanticipated: as COVID-19 was
beginning to take off in the US, some were already calling attention to the fact
that COVID-19 threatened to exacerbate existing disparities
(\protect\hyperlink{ref-kim_covid-19_2020}{Kim et al., 2020}). Kim, Marrast, and Conigliaro noted at least three structural
barriers in COVID-19 prevention and care: 1) originally requiring
residents to have a doctor's prescription for a COVID-19 test reduced the
opportunity for healthcare for People of Color as they are less likely to have a
primary care provider; 2) drive-through testing made testing disadvantaged those
who rely on public transportation; and 3) quarantining at home while waiting the
7-10 days originally required for COVID-19 test results to come back posed an
economic and social challenge that many in already financially difficult
situations may not have been able to take on (\protect\hyperlink{ref-kim_covid-19_2020}{2020}). Others noted
yet more reasons why COVID-19 threatened to worsen an already inequitable
healthcare landscape in the US: in particular, those who reside in prisons and jails,
immigrants and undocumented people, people with disabilities, and people
experiencing homeless all face additional challenges in seeking and getting the
healthcare they deserve (\protect\hyperlink{ref-okonkwo_covid-19_2021}{Okonkwo et al., 2021}). Even though stay-at-home orders
designed to mitigate spread that were prevalent in many states
(\protect\hyperlink{ref-moreland_timing_2020}{Moreland et al., 2020}), workers functioning in capacities essential to the
functioning of critical infrastructure operations (later termed ``essential
workers'') were exposed to heightened risk of COVID-19 transmission
(\protect\hyperlink{ref-hanage_covid-19_2020}{Hanage et al., 2020}; \protect\hyperlink{ref-national_bureau_of_economic_research_measuring_2021}{National Bureau of Economic Research, 2021}; \protect\hyperlink{ref-national_conference_of_state_legislatures_covid-19_2021}{National Conference of State Legislatures, 2021}; \protect\hyperlink{ref-the_lancet_plight_2020}{The Lancet, 2020}; \protect\hyperlink{ref-wei_risk_2022}{Wei et al., 2022}). It's clear that those with more
structurally enfranchised privileges have been more able to mitigate their risk
of negative health outcomes associated with COVID-19 --- during February 1 to
April 1, 2020, New York City residents who lived in more affluent neighborhoods
were more likely to have left the city, while New York City residents from more
Black and Hispanic
neighborhoods were more likely to continue working (\protect\hyperlink{ref-coven_disparities_2020}{Coven \& Gupta, 2020}).

\begin{quote}
Contrary to the oft used phrase that the `virus does not discriminate', the
data presented here suggest that this virus, as many other infectious diseases,
has the greatest implications for the most vulnerable people. The intersections
between health and human rights are clear---the health of a society and
vulnerability to a pandemic are directly related to its human rights track
record for those who are marginalised. (\protect\hyperlink{ref-okonkwo_covid-19_2021}{Okonkwo et al., 2021})
\end{quote}

When vaccines became available, vaccine appointments were often only available
to be scheduled through online web-portals contributing to the inequities
between those who had internet access and technological literacy and those who
didn't (\protect\hyperlink{ref-press_inequities_2021}{Press et al., 2021}). Vaccination sites have not been equally
distributed and areas determined to be vaccine deserts have been found to have
disproportionately more Black and Hispanic residents (\protect\hyperlink{ref-rader_spatial_2021}{Rader et al., 2021}). In
fact, healthcare facilities in counties with higher Black composition had 32\%
(95\% CI 14\%-47\%) lower odds of serving as vaccine sites
(\protect\hyperlink{ref-hernandez_disparities_2022}{Hernandez et al., 2022}).

What vaccination has been administered hasn't suddenly erased the unequal burden
of COVID-19 either; in August 2022 the New York Times was reporting ``Black death
rates at this winter's peak were greater than those of white people by 34
percent in rural areas, 40 percent in small or medium cities and 57 percent in
big cities and their suburbs'' (\protect\hyperlink{ref-goldstein_during_2022}{Goldstein, 2022}). As COVID-19 case and
mortality rates have waxed and waned, the inequities have widen and shrunken,
often with racial/ethnic inequities growing during times when COVID-19 rates
have surged (\protect\hyperlink{ref-hill_covid-19_2022}{Hill \& Artiga, 2022}). During 2022, the age-standardized
COVID-19 mortality rates for white people have, at times, been slightly
higher than those of Black and Hispanic people, predominantly because the
mortality rate among white people has increased (\protect\hyperlink{ref-johnson_whites_2022}{Johnson \& Keating, 2022}).
It's important to note that white COVID-19 mortality rates overtaking
the Black mortality rates does not imply that equity has been established: neither
does this undo the cumulative impact of mortality (which has been twice as high for
Black people compared to white people (\protect\hyperlink{ref-hill_covid-19_2022}{Hill \& Artiga, 2022})), nor does it imply
the underlying systemic barriers to equity have been overturned
(\protect\hyperlink{ref-del_rios_covid-19_2022}{Del Rios et al., 2022}). As Del Rios, Chomilo, and Lewis note, instead,
COVID-19 leaves in its wake more years of life expectancy lost, wages lost, and
degradation of mental health in Communities of Color.

\hypertarget{methods}{%
\section{Methods}\label{methods}}

\hypertarget{data-sources}{%
\subsection{Data Sources}\label{data-sources}}

The following variables were retrieved at the county level:

\begin{itemize}
\tightlist
\item
  Counts of COVID-19 deaths (\protect\hyperlink{ref-the_new_york_times_coronavirus_2021}{The New York Times, 2021}).
\item
  Population size estimates for 2020 from the U.S. Census (\protect\hyperlink{ref-us_census_bureau_2020_2021}{US Census Bureau, 2021}).
\item
  Median age, median household income, racial/ethnic composition, population
  density, percent below the federal poverty line, and number of households with
  high (\$100k+)/low (\textless\$25k) household income by racial/ethnic group from the
  2014-2019 5-year American Community Survey (\protect\hyperlink{ref-us_census_bureau_american_2020}{US Census Bureau, 2020})
  through the \texttt{tidyverse} R package (\protect\hyperlink{ref-walker_tidycensus_2022}{Walker \& Herman, 2022}).
\item
  Votes cast in the 2020 presidential election (\protect\hyperlink{ref-mit_election_data_and_science_lab_county_2022}{MIT Election Data and Science Lab, 2022})
\end{itemize}

\hypertarget{generalized-additive-models}{%
\subsection{Generalized Additive Models}\label{generalized-additive-models}}

Generalized additive models (GAMs) improve upon generalized linear models by
allowing for the fitting of smooth functions that transform right-hand-side variables.
This is a convenient means to account for nonlinear relationships between the
outcome and predictor variables. Whereas a generalized linear model may look like
\[ g(\mu_i) = \beta_0 + \beta_1 x_{i1} + \beta_2 x_{i2} + \beta_3 x_{i3} ...\]
a generalized additive model could look like
\[ g(\mu_i) = \mathbf A_i \mathbf \theta + f_1(x_{i1}) + f_2(x_{i2}) + f_3(x_{i3}, x_{i4}) + \dots\]
where the expected value of the outcome is given as \(\mu_i \equiv \mathbb E(Y_i)\),
\(\mathbf A_i\) is a row of the model matrix
for any strictly non-parametric model components, \(\mathbf \theta\) is the corresponding
parameter vector, and the \(f_j\) are smooth functions of the covariates \(x_k\) (\protect\hyperlink{ref-wood_generalized_2017}{S. N. Wood, 2017}).

\begin{figure}
\includegraphics[width=1\linewidth]{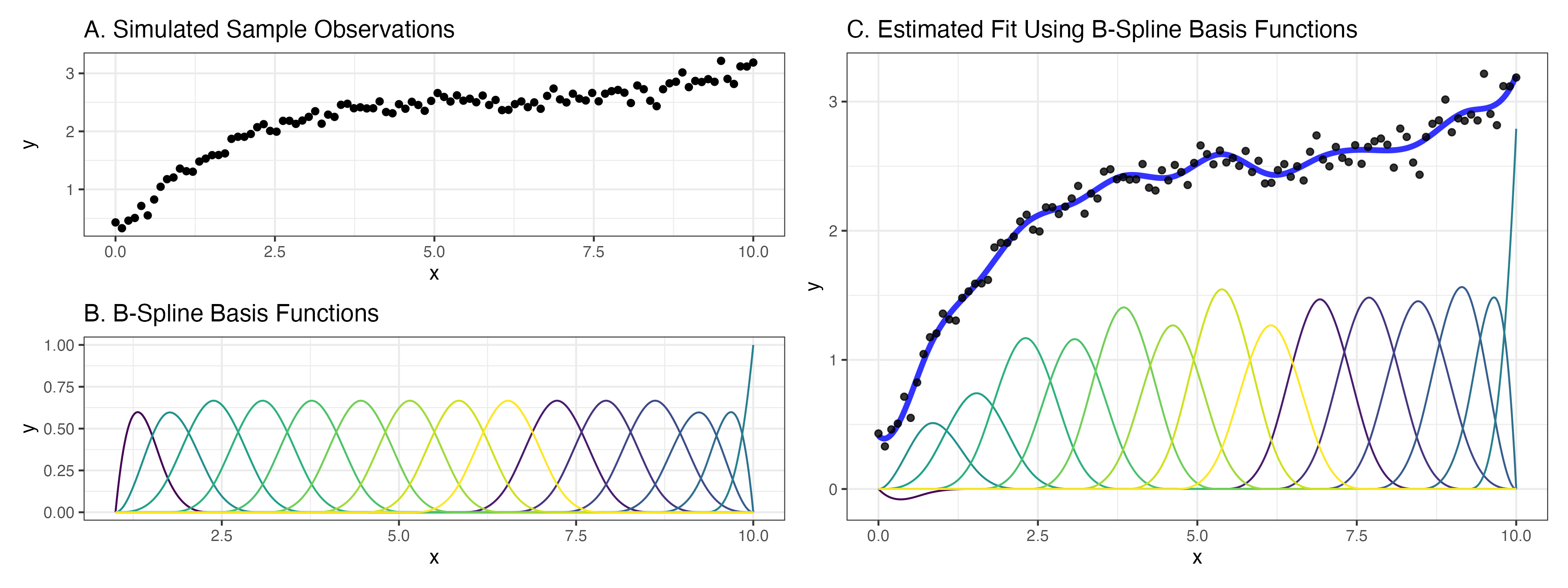} \caption{A demonstration of fitting regression models using B-spline basis functions. Part A. shows the simulated dataset. Part B. shows the B-spline basis functions. Part C. shows the weighting of the B-spline basis functions and their weighted sum with an intercept added to create the regression model fit to the data.}\label{fig:gam-demo-figure}
\end{figure}

The smooth functions estimated as part of fitting a GAM are constructed using
spline basis functions. These spline basis functions allow for the smooth
interpolation of trends in the data allowing for the incorporation of
nonlinearity. To avoid overfitting the data, a penalty term is introduced that
controls the degree of ``wiggliness'' or smoothness and this penalty term is fit
using generalized cross validation. B-splines are one kind of spline basis function
that is commonly used and are especially popular because of their property that they
are non-negative on only a finite interval.

While a variety of spline-based approaches exist (cubic splines, B- and
P-splines, thin-plate splines, etc.), our particular application setting
warrants the use of tensor-product smooths because we require anisotropic
penalization. That is to say, when we seek to create spatiotemporally smoothed
model estimates, it's inappropriate to assume that the amount of smoothing
across space should be the same as the amount of smoothing across time because
they're in fundamentally different units (space being measured in Cartesian
coordinates and time being measured in years, months, days, etc.). More details
about tensor-product smooths are available in Simon Wood's \emph{Generalized Additive Models - An Introduction with R} (\protect\hyperlink{ref-wood_generalized_2017}{2017}).

\hypertarget{overdispersion-and-event-modeling}{%
\subsection{Overdispersion and Event Modeling}\label{overdispersion-and-event-modeling}}

Above and beyond using the GAM framework to allow for flexible, nonlinear
relationships between our observed county-level variables and COVID-19
mortality, we must have a model specification that agrees with the data
generating process. In our case, the data generated are counts of deaths per
population, and the class of models most suited to represent counts of events
are Poisson, quasi-Poisson, and negative binomial models. Here we've chosen to
use the negative binomial model as it accounts for overdispersion and has
the intuitive interpretation of a Poisson model with gamma distributed underlying
rate parameter (\protect\hyperlink{ref-gelman_bayesian_2013}{Gelman et al., 2013}).

A negative binomial model in the context of modeling count data can be written
\[y_i \sim \text{NegativeBinomial}(u_i \theta_i, \phi),\]
where \(y_i\) are the counts observed, \(u_i\) is the ``exposure'', \(\theta_i\) are
the rates, and \(\phi\) determines the amount of overdispersion. The rates
are modeled as \(\theta_i = e^{X_i \beta}\) where \(X_i\) are the observed
covariates and \(\beta\) are the coefficients on the covariates corresponding
to log rate ratios. The logarithm of the exposure, \(\log(u_i)\) is often called
(and later herein referenced as) the offset. In epidemiological contexts, the offset
is often representative of the amount of person-time during which observations were recorded. Whereas the Poisson model holds that \(\text{var}(y) = \mu\) where \(\mu\)
is the average rate, the negative binomial model instead assumes that
\(\text{var}(y) = \mu + \mu^2/\phi\). The Poisson model
is a special case of the negative binomial model when \(\phi \to \infty\).

An alternative and equivalent formulation of the negative binomial that is commonly
used makes the connection to the Poisson model even more clear: instead of
\(\theta_i\) and \(\phi\), using \(\alpha\) and \(\beta\),
\[y \sim \text{NegativeBinomial}(\alpha, \beta), \text{ and}\]
\[\text{NegativeBinomial}(y|\alpha, \beta) = \int \text{Poisson}(y|\theta) \text{Gamma}(\theta | \alpha, \beta) d\theta.\]

Introductions to and additional exposition on the negative binomial model,
especially in the context of modeling counts of events outcome data, are
available in \emph{Bayesian Data Analysis} and \emph{Regression and Other Stories}
(\protect\hyperlink{ref-gelman_regression_2020}{Gelman et al., 2020}, \protect\hyperlink{ref-gelman_bayesian_2013}{2013}).

\hypertarget{variables-of-interest}{%
\subsection{Variables of Interest}\label{variables-of-interest}}

The following variables have been included as covariates of interest:

\begin{itemize}
\tightlist
\item
  Median Age
\item
  Population Density per Square Mile
\item
  Median Household Income
\item
  Proportion in Poverty
\item
  the Index of Concentration at the Extremes for Racialized Economic Segregation
\item
  Political Lean in the 2020 Election (1 = 100\% Republican votes, -1 = 100\% Democratic Votes)
\end{itemize}

\hypertarget{index-of-concentration-at-the-extremes-for-racialized-economic-segregation}{%
\paragraph{Index of Concentration at the Extremes for Racialized Economic Segregation}\label{index-of-concentration-at-the-extremes-for-racialized-economic-segregation}}

The Index of Concentration at the Extremes (ICE) is a measure which describes
how concentrated a given area's population is in terms of the extreme ends of
privilege and marginalization (\protect\hyperlink{ref-krieger_public_2016}{Krieger et al., 2016}).
In general, the ICE measure is formulated as

\[ \text{ICE} = \frac{\text{Number of People in Most Privileged Category} - \text{Number of People in Least Privileged Category}}{\text{Total Population}}\]

Applying the ICE approach to a specific context involves defining the axes of
privilege of interest. In this case, data on racialized economic segregation
are used from the US Census American Community Survey on white households
earning more than \$100,000 a year (the most privileged group) or households of
People of Color earning less than \$25,000 a year (the least privileged group).
This variable is referred to throughout as the ICE for racialized economic
segregation, or \texttt{ICEraceinc} in the code.

Compared with the Gini
coefficient which is one of the most popular methods for summarizing area-level
rates of inequities, the ICE has the advantage that it is suitable for
describing inequities at smaller area levels (\protect\hyperlink{ref-krieger_public_2016}{Krieger et al., 2016}). While the
Gini coefficient measures within-area dissimilarity (as in, for example, how
unequal wealth is distributed within a county), the ICE measure establishes
where on a spectrum a given county's population falls allowing for comparison
across counties. The Gini coefficient suffers from the fact that areas which are
made up of relatively homogeneous populations will appear as having low
within-area inequality (and therefore low Gini coefficient). Instead, the ICE
measures how much of the population is either privileged or not. The Gini
coefficient remains useful for reporting on the degree of inequity in larger
areal units (like countries, states, and regions), but at smaller areal units
(like counties, ZIP codes, census tracts) can be more difficult to interpret and
compare. Maps of ICE measure can elucidate what spatial social segregation and
polarization exist, and the ICE for racialized economic segregation has been
repeatedly and significantly associated with COVID-19 outcomes
(\protect\hyperlink{ref-brown_ecological_2021}{Brown et al., 2021}; \protect\hyperlink{ref-chen_revealing_2021}{Chen \& Krieger, 2021}; \protect\hyperlink{ref-eichenbaum_health_2022}{Eichenbaum \& Tate, 2022}; \protect\hyperlink{ref-hanage_covid-19_2020}{Hanage et al., 2020}; \protect\hyperlink{ref-krieger_relationship_2022}{Krieger et al., 2022}; \protect\hyperlink{ref-saha_neighborhood-level_2020}{Saha \& Feldman, 2020}; \protect\hyperlink{ref-sonderlund_racialized_2022}{Sonderlund et al., 2022}).

\hypertarget{political-lean}{%
\paragraph{Political Lean}\label{political-lean}}

Political lean has been associated with COVID-19 mortality in numerous studies,
with plausible mechanisms explaining the association including differences in
non-pharmaceutical intervention uptake (mask usage, social distancing,
quarantining), differences in rates of vaccination, differences in political
leadership's messaging, resource allocation, and the adoption of policy
interventions (\protect\hyperlink{ref-gonzalez_conservatism_2021}{Gonzalez et al., 2021}; \protect\hyperlink{ref-grossman_political_2020}{Grossman et al., 2020}; \protect\hyperlink{ref-kaashoek_evolving_2021}{Kaashoek et al., 2021}; \protect\hyperlink{ref-krieger_relationship_2022}{Krieger et al., 2022}; \protect\hyperlink{ref-leonhardt_red_2021}{Leonhardt, 2021}).

\hypertarget{results}{%
\section{Results}\label{results}}

\hypertarget{application-1-non-spatial-covariate-effects-over-time}{%
\subsection{Application 1: Non-Spatial Covariate Effects Over Time}\label{application-1-non-spatial-covariate-effects-over-time}}

\begin{figure}

{\centering \includegraphics[width=1\linewidth,height=0.75\textheight]{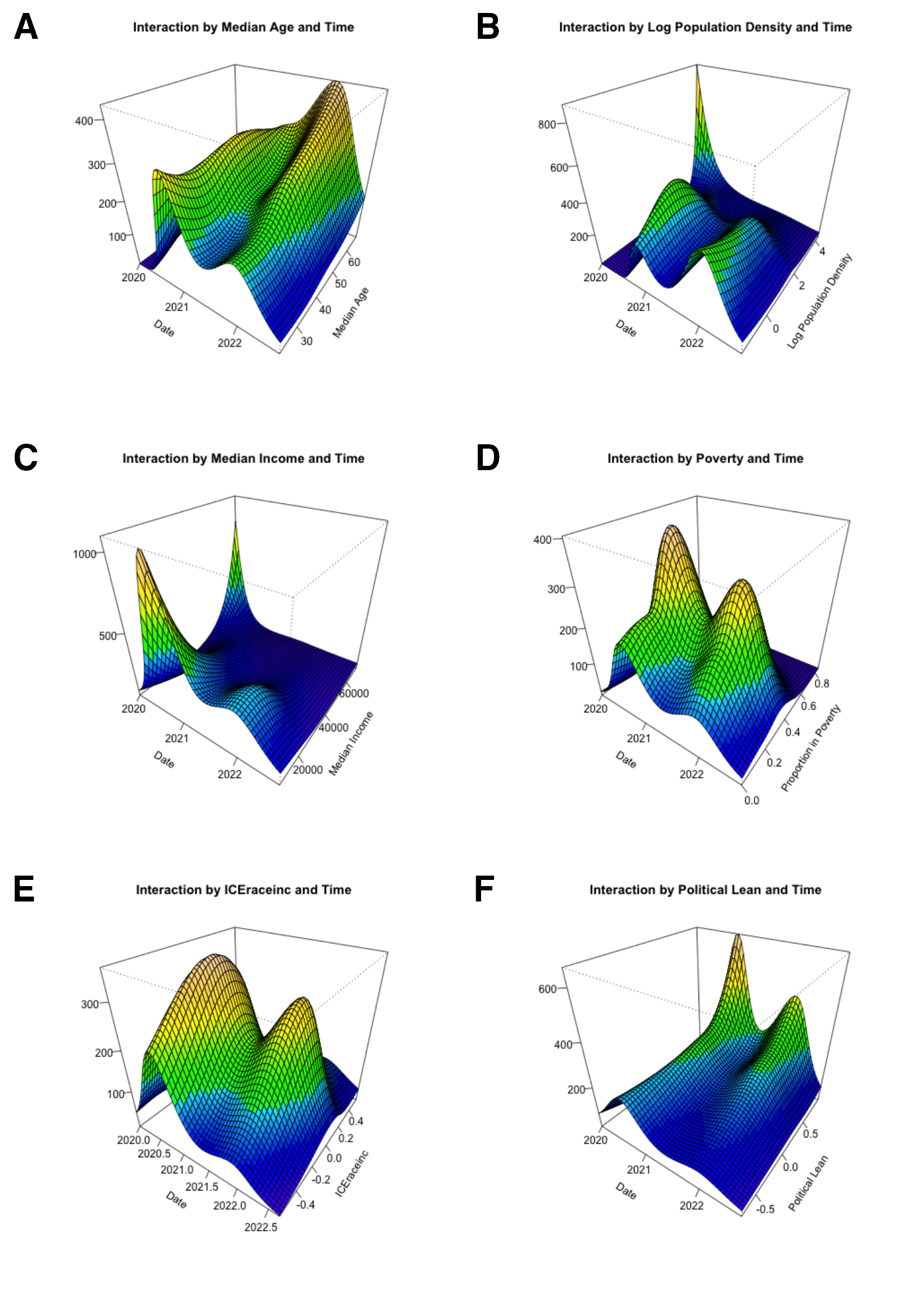}

}

\caption{Additional COVID-19 mortality associated with covariates over time. A) age and time, B) log (base 10) population density and time, C) median income and time, D) proportion in poverty and time, E) ICEraceinc and time, F) political lean and time. In panels B-F the effect of age is marginalized out using the median age in the US, 38.8 (\protect\hyperlink{ref-us_census_bureau_nation_2022}{US Census Bureau, 2022}).}\label{fig:one-variable-over-time}
\end{figure}

The GAM models shown in Figure \ref{fig:one-variable-over-time} are fit with the following structure using
the \texttt{gam} function from the \texttt{mgcv} package (\protect\hyperlink{ref-wood_mgcv_2022}{S. Wood, 2022}):

\begin{Shaded}
\begin{Highlighting}[]
\FunctionTok{gam}\NormalTok{(}
  \AttributeTok{formula =}\NormalTok{ deaths }\SpecialCharTok{\textasciitilde{}} \FunctionTok{s}\NormalTok{(median\_age) }\SpecialCharTok{+} \FunctionTok{te}\NormalTok{(covariate, date, }\AttributeTok{d=}\FunctionTok{c}\NormalTok{(}\DecValTok{1}\NormalTok{,}\DecValTok{1}\NormalTok{)), }\CommentTok{\# regression formula}
  \AttributeTok{offset =} \FunctionTok{log}\NormalTok{(popsize}\SpecialCharTok{/}\FloatTok{1e5}\SpecialCharTok{/}\DecValTok{12}\NormalTok{), }\CommentTok{\# our offset represents the person{-}time}
  \AttributeTok{data =}\NormalTok{ df,     }\CommentTok{\# our data{-}set of county{-}level observations}
  \AttributeTok{family =} \FunctionTok{nb}\NormalTok{(}\AttributeTok{link=}\StringTok{\textquotesingle{}log\textquotesingle{}}\NormalTok{)  }\CommentTok{\# indicates negative binomial family and a log{-}link function}
\NormalTok{)}
\end{Highlighting}
\end{Shaded}

The formula used puts a one-dimensional smoothing spline on median age to represent a
nonlinear age-effect and a two-dimensional tensor-product smooth on the
interaction between the given covariate and the date. The \texttt{d=c(1,1)} argument
provides the instruction necessary to consider covariate and date as being on
separate scales and therefore to fit the tensor-product smooth with anisotropy ---
that is, to allow for independent amounts of scaling in the dimensions of the
covariate and time. The offset used structures the regression to estimate rates
in units of person-time per 100,000 person-years. Since the death counts are
aggregated to the monthly level, the person-time in units of 100,000
person-years are calculated by taking each county's population size, dividing by
100,000, and dividing by 12 (for the 12 months in a year).

The above structure is used to estimate models for our different \texttt{covariate}
variables of interest: median income, percent in poverty, the ICE for racialized
economic segregation, political lean. The model presenting median age treats
median age as the main covariate of interest including it as the \texttt{te(covariate,\ date,\ d=c(1,1))} and dropping the \texttt{s(median\_age)} term which otherwise becomes redundant.
Results of these models are summarized in Figure
\ref{fig:one-variable-over-time} where the additional COVID-19 mortality
associated with each covariate is visualized.

Likelihood ratio tests confirmed that models with covariates interacted with
time had significantly lower residual deviance (\(p \leq 2.2e^{-16}\) for all
models) compared to models only including a spline term on median age and the
given covariate not interacted with time. The model interacting median age
and time was compared to a model with a spline for median age alone. Akaike Information
Criteria values were also lower for all models compared to models that did
not interact the covariates with time.

\begin{figure}
\includegraphics[width=1\linewidth]{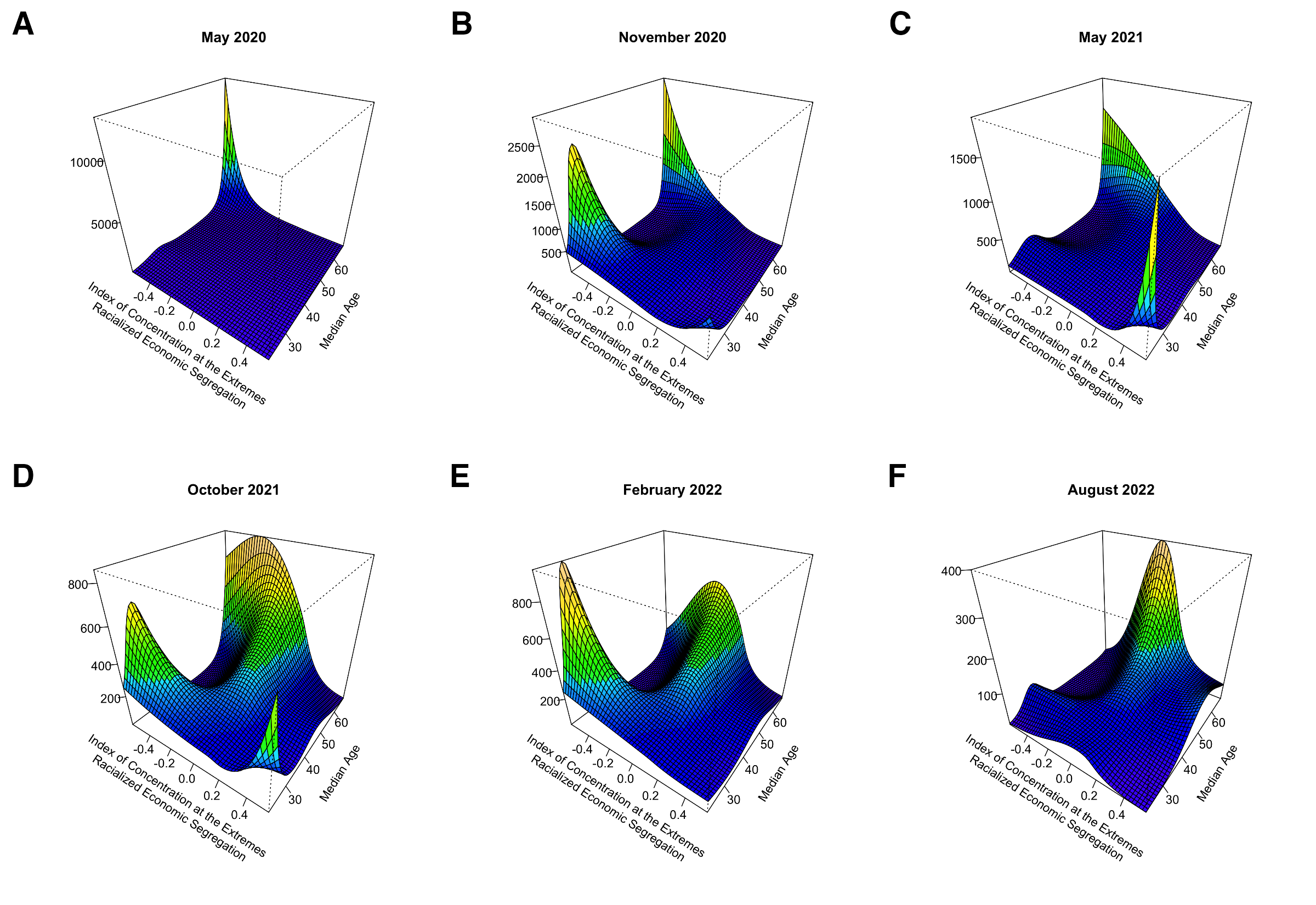} \caption{The changing interacted effect of the ICE for racialized economic segregation and age over time. A) May 2020, B) November 2020, C) May 2021, D) October 2021, E) February 2022, F) August 2022. An animated version is available online at \url{https://github.com/ctesta01/covid.gradient.estimation/blob/main/analysis/05_two_variables_at_a_time/animation_ICEraceinc_age/readme.md}}\label{fig:two-variables-over-time}
\end{figure}

A more complex non-spatial application of GAMs to describe the distribution of
COVID-19 mortality in the US over time is to consider the associations of
mortality with three-way interactions of time and two covariates taken together.
In the following example, the formula used is \texttt{deaths\ \textasciitilde{}\ te(date,\ ICEraceinc,\ median\_age,\ d=c(1,1,1))}. Again, the \texttt{d} argument specifying the marginal
basis dimensions is used indicate that each of the \texttt{date}, \texttt{ICEraceinc}, and
\texttt{median\_age} measures are in different units and should not be smoothed assuming
that a one unit difference in one variable is comparable to a one unit difference
in another variable. This approach is represented in Figure \ref{fig:two-variables-over-time}.

\newpage

\hypertarget{application-2-spatiotemporal-smoothing}{%
\subsection{Application 2: Spatiotemporal Smoothing}\label{application-2-spatiotemporal-smoothing}}

By fitting GAMs with a tensor-product smoothing term on latitude, longitude, and
time we can estimate a spatiotemporally smoothed trend in COVID-19 mortality. To
do this, the GAM is constructed similarly as in Application 1 but with the
smoothing term specified as \texttt{te(latitude,\ longitude,\ time,\ d=c(2,1))} where
\texttt{d=c(2,1)} indicates that \texttt{latitude} and \texttt{longitude} share the same dimensions
(i.e., both are spatial and in units of degrees) while the \texttt{time} data are in
separate units. Note that results presented in Application 2 and 3 are based on
county-level data from the contiguous US excluding Alaska and Hawaii.

Using GAMs to present spatiotemporally smoothed estimates of COVID-19 mortality
allows for the synthesis of trends over time. Whereas the raw rates of COVID-19
mortality are noisy due to 72\% of the land area in the US being classified as
rural and low population density (\protect\hyperlink{ref-health_resources__services_administration_defining_2022}{Health Resources \& Services Administration, 2022}), the spatiotemporally smoothed GAM estimates
use population weighting via the offset specified to estimate a latent surface
that represents localized mortality rate averages in spatial coordinates and in
time.

\begin{figure}
\includegraphics[width=1\linewidth]{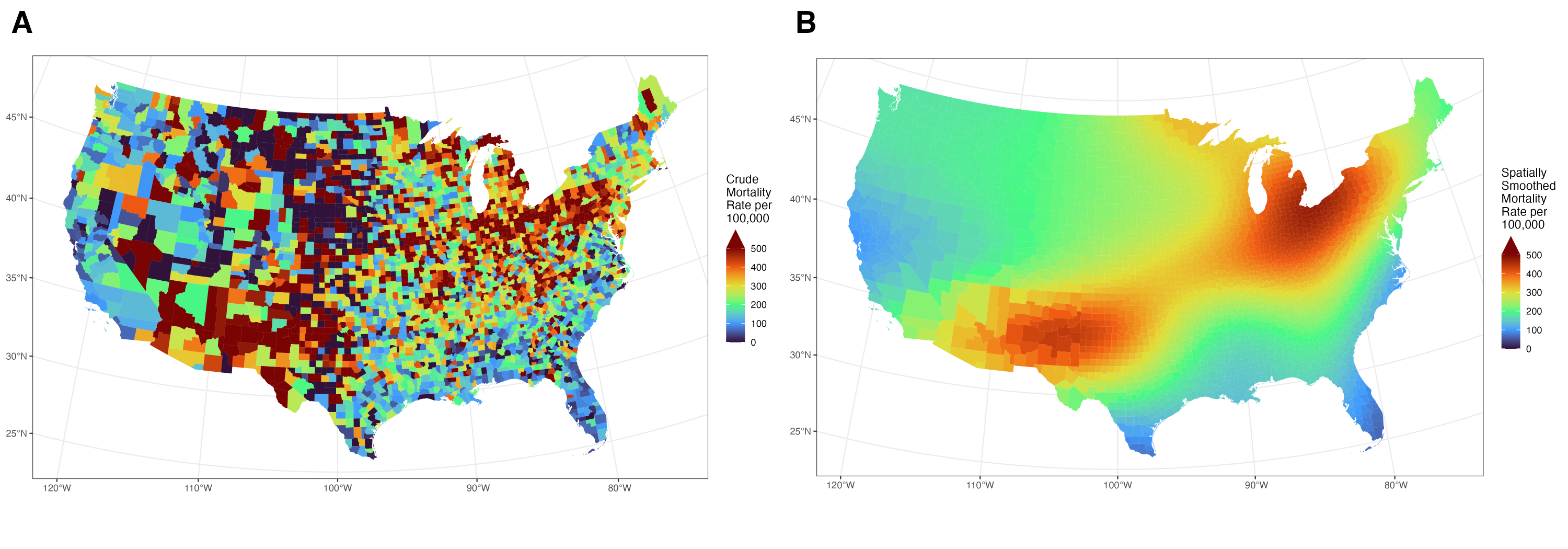} \caption{Comparison of crude vs. smoothed mortality rates. A) Crude mortality rates per 100,000 person-years in January 2022. B) Smoothed mortality rates from a GAM applied to mortality rates from January 2022.}\label{fig:crude-vs-smoothed}
\end{figure}

Figure \ref{fig:crude-vs-smoothed} shows the difference between crude mortality
rates and spatially smoothed mortality rates in January 2022 to illustrate the
level of noise present in raw rates and how spatially smoothing synthesizes
local geographic patterns into trends that can be meaningfully
interpreted as local area mortality risk levels with information information pooled across
nearby county rates.

\begin{figure}
\includegraphics[width=1\linewidth]{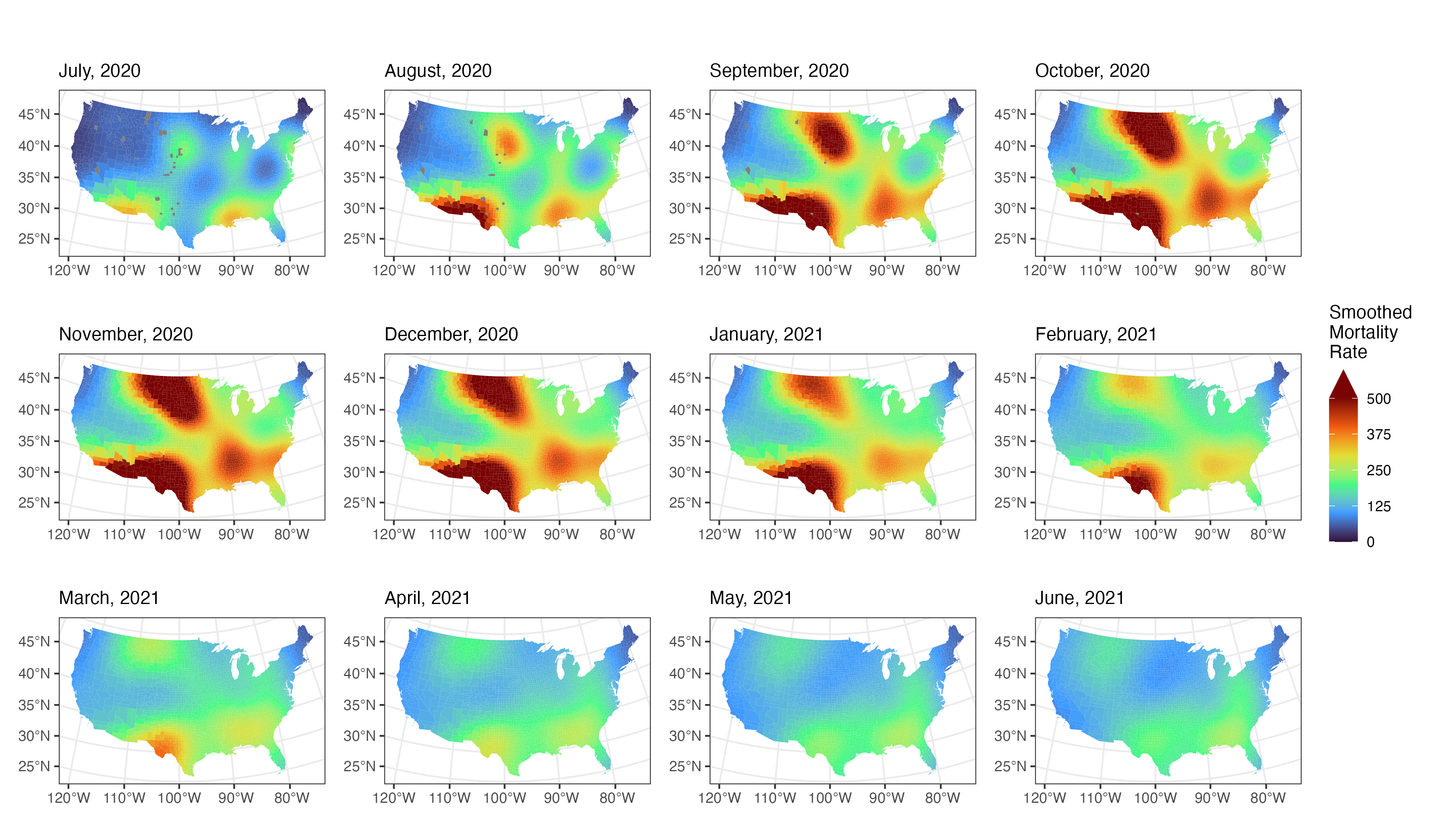} \caption{Spatiotemporally smoothed COVID-19 mortality estimates from a GAM fit to data from March 2020-August 2022. Panels show months selected to highlight the changing spatial patterns of COVID-19 mortality risk over time. An animated version is available online at \url{https://github.com/ctesta01/covid.gradient.estimation/blob/main/analysis/09_spatiotemporal_models/animation/readme.md}}\label{fig:spatiotemporal-panel-figure}
\end{figure}

Figure \ref{fig:spatiotemporal-panel-figure} shows the results from a
spatiotemporally smoothed model in select months to highlight how spatiotemporal
smoothing can yield results that synthesize trends in space and
time.

\newpage

\hypertarget{application-3-estimating-covariate-effects-adjusted-for-spatiotemporal-autocorrelation}{%
\subsection{Application 3: Estimating Covariate Effects Adjusted for Spatiotemporal Autocorrelation}\label{application-3-estimating-covariate-effects-adjusted-for-spatiotemporal-autocorrelation}}

The final application of GAMs presented here is to estimate the effects of
covariates after adjusting for a spatiotemporally smoothed latent risk surface.
This has the interpretation of asking what are the changes to
COVID-19 mortality rates associated with each covariate after taking into
account regional geographic patterns in COVID-19 mortality over time. In
particular, this is relevant for understanding what county-level measures are
associated with elevated COVID-19 mortality even after adjusting for where
COVID-19 rates were locally elevated or depressed due to variation in the timing
of local introduction, transmission, and diffusion.

\begin{figure}
\includegraphics[width=1\linewidth]{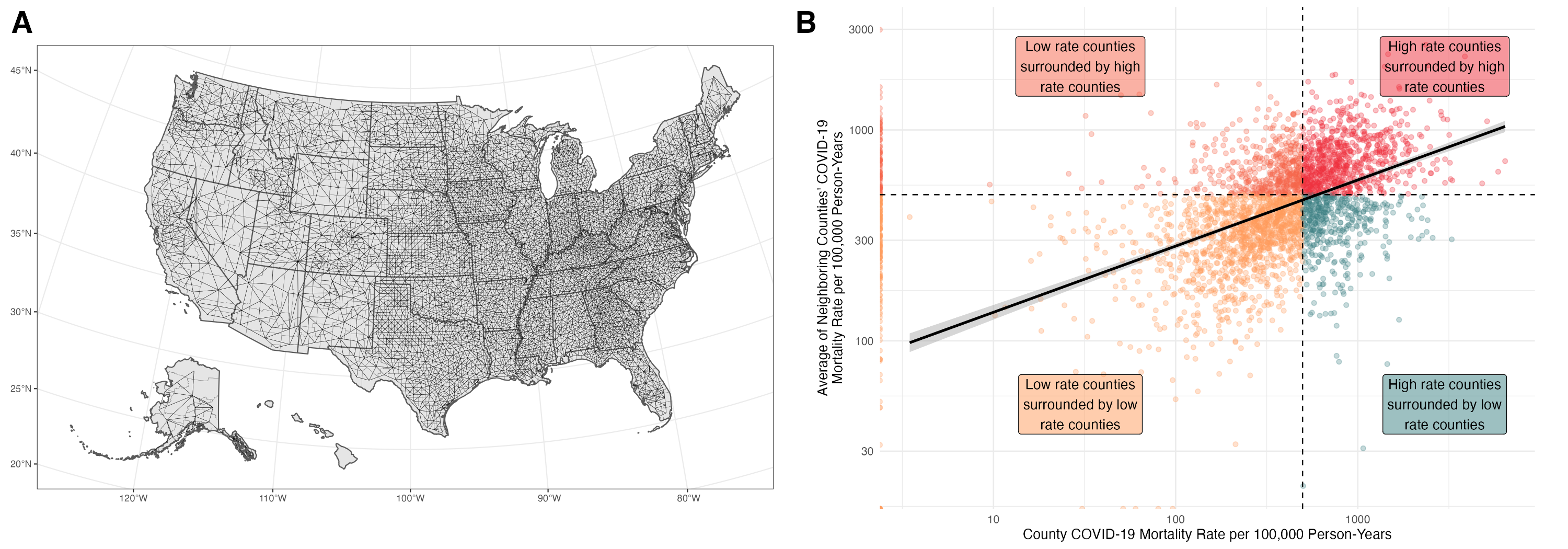} \caption{The spatial autocorrelation of COVID-19 rates in January 2022. Panel A shows the graph of US counties connected according to which are neighbors. Panel B shows a scatterplot depicting each US counties' COVID-19 mortality rate (x-axis) compared to the average of its neighbors' COVID-19 mortality rates (y-axis). A regression line shows the association between counties' COVID-19 rates and their neighboring counties' COVID-19 mortality rates.  Dotted lines indicated the average for US county-level crude COVID-19 mortality rates (vertical) and for the average of each counties' neighboring counties' COVID-19 mortality rates (horizontal).}\label{fig:neighboring-graph-morans-i}
\end{figure}

\begin{figure}

{\centering \includegraphics[width=0.5\linewidth]{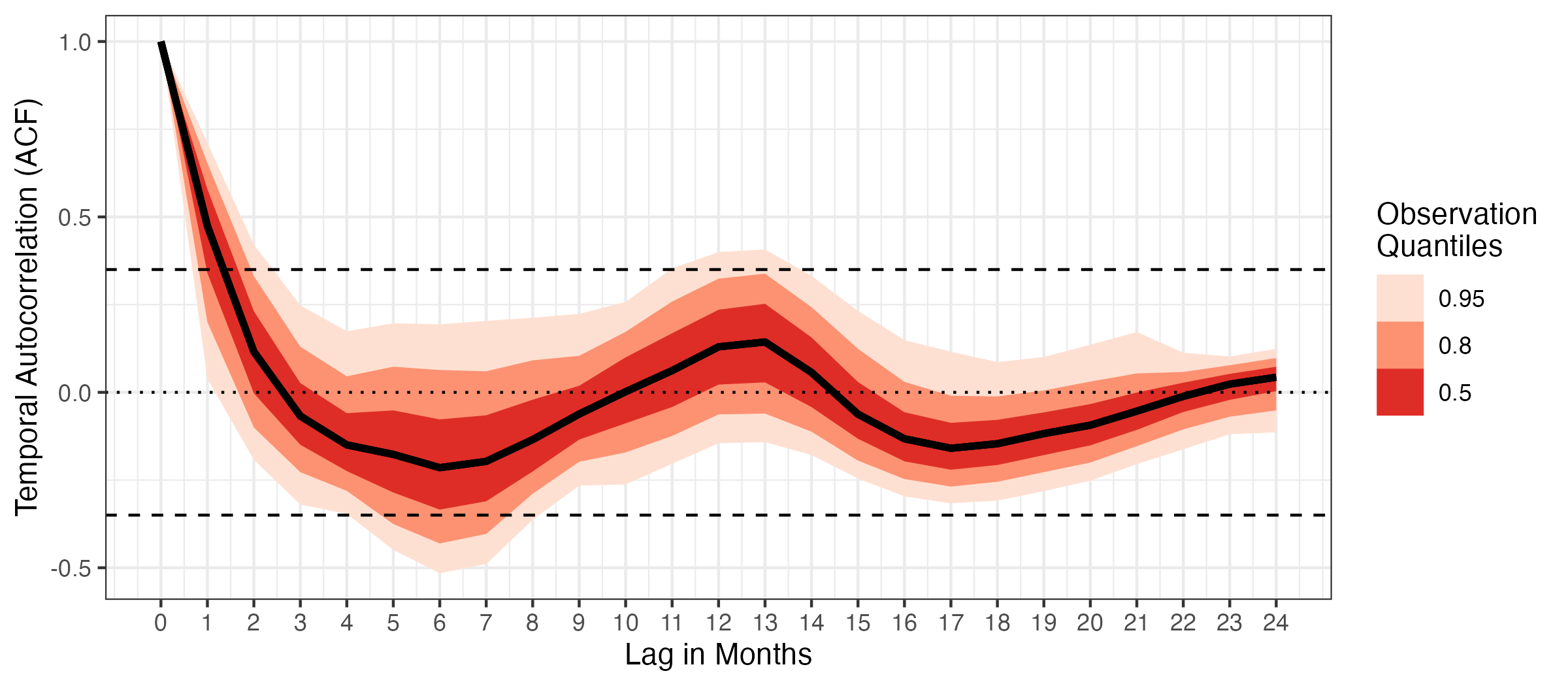}

}

\caption{The median (black) and quantiles (shaded bands) of temporal autocorrelation of counties' COVID-19 mortality rates over time during April 2020 — August 2022.}\label{fig:temporal-acf}
\end{figure}

To motivate the need for and application of spatiotemporal smoothing in
assessing the associated changes in COVID-19 mortality with county-level
covariates, Figure \ref{fig:neighboring-graph-morans-i} presents the Moran's I
diagnostic plot for January 2022 which summarizes the amount of autocorrelation
between counties' COVID-19 mortality rates and the average of
each counties' neighboring counties' COVID-19 mortality rates. The observed
autocorrelation indicates that there is an association between counties'
COVID-19 rates and their neighboring counties' COVID-19 mortality rates, implying that
without taking this patterning into account, regression models that do not
explicitly model the effect of spatial autocorrelation may be biased due to
inappropriately assuming that county-level data are independent of one another. Second,
while the Moran's I plot demonstrates the spatial-correlation present in the data
at a single time-point, the following Figure \ref{fig:temporal-acf} shows the
tendency of county COVID-19 observations to be autocorrelated between subsequent
months.

\begin{figure}

{\centering \includegraphics[width=0.85\linewidth]{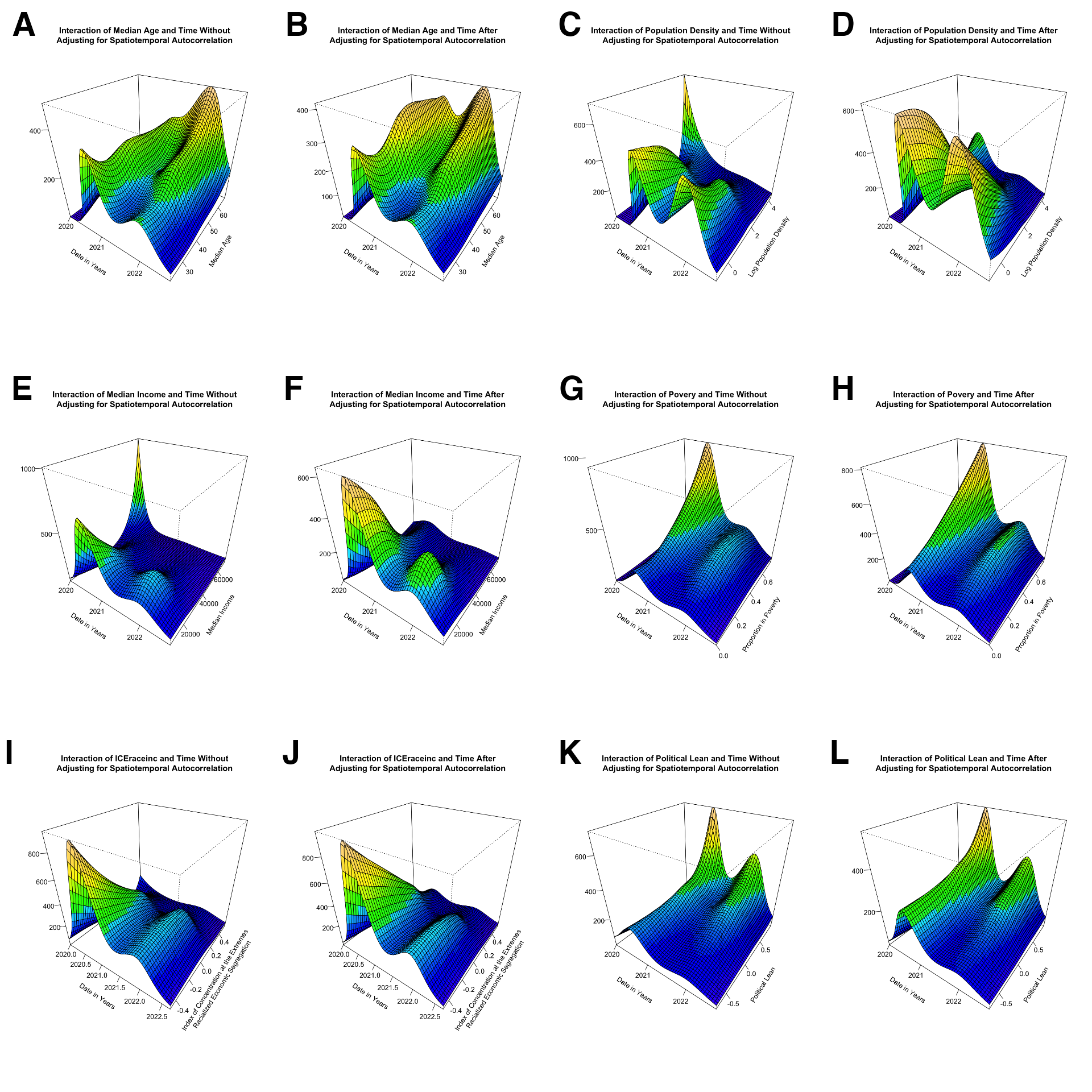}

}

\caption{Estimates of increases in COVID-19 mortality associated with county-level covariates without (first) and with (second) adjustment for spatiotemporal autocorrelation. Panels A and B show COVID-19 mortality increases associated with varying levels of median age over time; C and D show associations with log population density; E and F median income; G and H for proportion in poverty; I and J the ICE for racialized economic segregation; K and L political lean.}\label{fig:with-and-without-spatiotemporal-adjustment}
\end{figure}

Figure \ref{fig:with-and-without-spatiotemporal-adjustment} shows the
additional COVID-19 mortality associated with each covariate over time without and
with adjustment for spatiotemporal trends. The models for population density
and median income show particularly marked changes with the spike at high
population density and high median incomes disappearing after adjusting for a
spatiotemporally smoothed term. This likely reflects the early emergence and
surge of COVID-19 in the greater New York City area as being predominantly a
feature of the combined effect of human geography and infectious disease
dynamics where the earliest introductions of COVID-19 were made to dense social
networks in the US. Other effects remain qualitatively changed little,
suggesting that the trends in COVID-19 mortality associated with these
county-level measures remains durable when taking into account spatiotemporal
autocorrelation.

\newpage

\hypertarget{discussion}{%
\section{Discussion}\label{discussion}}

This paper presents three applications of GAMs to county-level COVID-19
mortality data from the US during March 2020 to August 2022. This paper
addresses an important need, which is to assess what measures of the
socioeconomic and geopolitical landscape are associated with increases in
COVID-19 mortality in a way that takes into account how infectious disease
dynamics are locally correlated in space and time. However, the approaches
outlined in this paper are not without several limitations which include (at
least) the following: the potential for cross-level confounding, lack of
individual level data, an inability to include all possible metrics, further
improvements to the methods used may be warranted, and an inability to make
claims about the causal relationships between the variables investigated.

In analyses that use area-level aggregated data, the ecologic fallacy refers to
the possibility that associations between area-level measures and outcomes may
not reflect the associations that would be observed if analyzing individual
level data on the same measures and outcomes as a result of cross-level
confounding (\protect\hyperlink{ref-greenland_ecologic_2001}{Greenland, 2001}). It's important to note that
some of the variables in this analysis are included to contextualize the places
in which people live, reside, and work and area-based social metrics should be
included even in individual-level analyses so as to accurately capture the
effects of both individual-level risk factors and risk factors that stem from
the contexts in which people exist (\protect\hyperlink{ref-testa_public_2022}{Testa et al., 2022}).

One particular improvement that could be made in future work is to address age
adjustment by using direct or indirect age standardization (\protect\hyperlink{ref-anderson_age_1998}{Anderson et al., 1998}) or
to create age-group stratified models. In the applications presented here,
county-level median age is adjusted for in order to control for age effects as
age-specific data were not publicly available without substantial levels of
suppression.

Additional variables like vaccination rates, mobility data, further variables
that relate to risk factors for COVID-19, and other COVID-19 related outcomes
like case trends and hospitalizations may be worthwhile to investigate under a
similar framework. The emphasis on trends in COVID-19 mortality in the
applications in this paper was done to prioritize presenting data that are less
subject to the reporting inconsistencies in COVID-19 cases and
hospitalizations (\protect\hyperlink{ref-galaitsi_challenges_2021}{Galaitsi et al., 2021}).

Further improvements to the methods employed here could include use of soap film
smoothing splines which are more appropriate in settings where non-convex
geographic boundaries are common (as in with peninsulas and bays)
(\protect\hyperlink{ref-wood_soap_2008}{S. N. Wood et al., 2008}). Taking a critical eye to the assumptions employed reveals yet
another area of potential improvement, which is the nature of how\\
smoothing splines are penalized in the GAM framework. It is conceivable that
geographic changes in COVID-19 mortality rates are not equally smooth across
all regions of the US, potentially warranting rapid changes in some areas
and more smooth changes in other areas in a risk surface model. The Bayesian
wombling literature provides the means to identify areas of rapid change in
latent surface models and may pose a fruitful direction of exploration for
future efforts to understand trends in COVID-19 mortality rates in the US
(\protect\hyperlink{ref-gelfand_bayesian_2015}{Gelfand \& Banerjee, 2015}).

\hypertarget{conclusions}{%
\section{Conclusions}\label{conclusions}}

This work introduces generalized additive models (GAMs) applied to COVID-19
mortality data to document the disparities and inequities in which US counties
suffered the worst mortality outcomes and at what times. The effort to document
inequities and explain the driving mechanisms behind them is crucial in building
up the evidence necessary to enact policy changes that can mitigate unjust and
preventable harm in the future. To this end, it is critical that efforts to
model infectious disease dynamics take into account how disease transmission and
outcomes are spatiotemporally patterned, violating any misplaced assumptions
that area-level disease rates can be treated as independent observations. By
presenting novel applications of GAMs to trends in COVID-19 mortality, this
paper illustrates new ways of understanding and visualizing the disproportionate
burden some communities have suffered by incorporating complex patterns of
interaction across socioeconomic and geopolitical variables and spatiotemporal
trends.

\newpage

\hypertarget{references}{%
\section*{References}\label{references}}
\addcontentsline{toc}{section}{References}

\hypertarget{refs}{}
\begin{CSLReferences}{1}{0}
\leavevmode\vadjust pre{\hypertarget{ref-anderson_age_1998}{}}%
Anderson, R. N., Rosenberg, H. M., \& National Center for Health Statistics. (1998). Age standardization of death rates; implementation of the year 2000 standard. \emph{National Vital Statistics Report}. \url{https://stacks.cdc.gov/view/cdc/13357}

\leavevmode\vadjust pre{\hypertarget{ref-bailey_how_2021}{}}%
Bailey, Z. D., Feldman, J. M., \& Bassett, M. T. (2021). How {Structural} {Racism} {Works} --- {Racist} {Policies} as a {Root} {Cause} of {U}.{S}. {Racial} {Health} {Inequities}. \emph{New England Journal of Medicine}, \emph{384}(8), 768--773. \url{https://doi.org/10.1056/NEJMms2025396}

\leavevmode\vadjust pre{\hypertarget{ref-bendix_less_2022}{}}%
Bendix, A. (2022). Less than 4\% of eligible people have gotten updated {Covid} booster shots, one month into the rollout. In \emph{NBC News}. \url{https://www.nbcnews.com/health/health-news/updated-covid-booster-shots-doses-administered-cdc-rcna48960}

\leavevmode\vadjust pre{\hypertarget{ref-blendon_inequities_2002}{}}%
Blendon, R. J., Schoen, C., DesRoches, C. M., Osborn, R., Scoles, K. L., \& Zapert, K. (2002). Inequities {In} {Health} {Care}: {A} {Five}-{Country} {Survey}. \emph{Health Affairs}, \emph{21}(3), 182--191. \url{https://doi.org/10.1377/hlthaff.21.3.182}

\leavevmode\vadjust pre{\hypertarget{ref-brown_ecological_2021}{}}%
Brown, K. M., Lewis, J. Y., \& Davis, S. K. (2021). An ecological study of the association between neighborhood racial and economic residential segregation with {COVID}-19 vulnerability in the {United} {States}' capital city. \emph{Annals of Epidemiology}, \emph{59}, 33--36. \url{https://doi.org/10.1016/j.annepidem.2021.04.003}

\leavevmode\vadjust pre{\hypertarget{ref-carpenter_health_2021}{}}%
Carpenter, E. (2021). {``{The} {Health} {System} {Just} {Wasn}'t {Built} for {Us}''}: {Queer} {Cisgender} {Women} and {Gender} {Expansive} {Individuals}' {Strategies} for {Navigating} {Reproductive} {Health} {Care}. \emph{Women's Health Issues}, \emph{31}(5), 478--484. \url{https://doi.org/10.1016/j.whi.2021.06.004}

\leavevmode\vadjust pre{\hypertarget{ref-cdc_location_2021}{}}%
CDC. (2021). Location, location, location --~{COVID} {Data} {Tracker} {Weekly} {Review}. In \emph{Centers for Disease Control and Prevention}. \url{https://www.cdc.gov/coronavirus/2019-ncov/covid-data/covidview/past-reports/07162021.html}

\leavevmode\vadjust pre{\hypertarget{ref-chen_revealing_2021}{}}%
Chen, J. T., \& Krieger, N. (2021). Revealing the {Unequal} {Burden} of {COVID}-19 by {Income}, {Race}/{Ethnicity}, and {Household} {Crowding}: {US} {County} {Versus} {Zip} {Code} {Analyses}. \emph{Journal of Public Health Management and Practice}, \emph{27}(1), S43--S56. \url{https://doi.org/10.1097/PHH.0000000000001263}

\leavevmode\vadjust pre{\hypertarget{ref-chrisler_ageism_2016}{}}%
Chrisler, J. C., Barney, A., \& Palatino, B. (2016). Ageism can be {Hazardous} to {Women}'s {Health}: {Ageism}, {Sexism}, and {Stereotypes} of {Older} {Women} in the {Healthcare} {System}. \emph{Journal of Social Issues}, \emph{72}(1), 86--104. \url{https://doi.org/10.1111/josi.12157}

\leavevmode\vadjust pre{\hypertarget{ref-coven_disparities_2020}{}}%
Coven, J., \& Gupta, A. (2020). \emph{Disparities in {Mobility} {Responses} to {COVID}-19}. \url{https://arpitgupta.info/s/DemographicCovid.pdf}

\leavevmode\vadjust pre{\hypertarget{ref-del_rios_covid-19_2022}{}}%
Del Rios, M., Chomilo, N. T., \& Lewis, N. A. (2022). Covid-19 is an inverse equity story, not a racial equity success story. In \emph{STAT}. \url{https://www.statnews.com/2022/10/25/covid-19-inverse-equity-story-not-racial-equity-success-story/}

\leavevmode\vadjust pre{\hypertarget{ref-donovan_us_2022}{}}%
Donovan, D. (2022). U.{S}. {Officially} {Surpasses} 1 {Million} {COVID}-19 {Deaths}. In \emph{Johns Hopkins Coronavirus Resource Center}. \url{https://coronavirus.jhu.edu/from-our-experts/u-s-officially-surpasses-1-million-covid-19-deaths}

\leavevmode\vadjust pre{\hypertarget{ref-eichenbaum_health_2022}{}}%
Eichenbaum, A., \& Tate, A. D. (2022). Health {Inequity} in {Georgia} {During} the {COVID}-19 {Pandemic}: {An} {Ecological} {Analysis} {Assessing} the {Relationship} {Between} {County}-{Level} {Racial}/{Ethnic} and {Economic} {Polarization} {Using} the {ICE} and {SARS}-{CoV}-2 {Cases}, {Hospitalizations}, and {Deaths} in {Georgia} as of {October} 2020. \emph{Health Equity}, \emph{6}(1), 230--239. \url{https://doi.org/10.1089/heq.2021.0118}

\leavevmode\vadjust pre{\hypertarget{ref-feldman_health_2021}{}}%
Feldman, J. L., Luhur, W. E., Herman, J. L., Poteat, T., \& Meyer, I. H. (2021). Health and health care access in the {US} transgender population health ({TransPop}) survey. \emph{Andrology}, \emph{9}(6), 1707--1718. \url{https://doi.org/10.1111/andr.13052}

\leavevmode\vadjust pre{\hypertarget{ref-galaitsi_challenges_2021}{}}%
Galaitsi, S. E., Cegan, J. C., Volk, K., Joyner, M., Trump, B. D., \& Linkov, I. (2021). The challenges of data usage for the {United} {States}' {COVID}-19 response. \emph{International Journal of Information Management}, \emph{59}, 102352. \url{https://doi.org/10.1016/j.ijinfomgt.2021.102352}

\leavevmode\vadjust pre{\hypertarget{ref-gelfand_bayesian_2015}{}}%
Gelfand, A. E., \& Banerjee, S. (2015). Bayesian wombling: Finding rapid change in spatial maps. \emph{WIREs Computational Statistics}, \emph{7}(5), 307--315. \url{https://doi.org/10.1002/wics.1360}

\leavevmode\vadjust pre{\hypertarget{ref-gelman_bayesian_2013}{}}%
Gelman, A., Carlin, J. B., Stern, H. S., Dunson, D. B., Vehtari, A., \& Rubin, D. B. (2013). Bayesian {Data} {Analysis}. In \emph{Routledge \& CRC Press}. \url{https://www.routledge.com/Bayesian-Data-Analysis/Gelman-Carlin-Stern-Dunson-Vehtari-Rubin/p/book/9781439840955}

\leavevmode\vadjust pre{\hypertarget{ref-gelman_regression_2020}{}}%
Gelman, A., Hill, J., \& Vehtari, A. (2020). \emph{Regression and {Other} {Stories}} (1st ed.). Cambridge University Press. \url{https://doi.org/10.1017/9781139161879}

\leavevmode\vadjust pre{\hypertarget{ref-glenza_covid_2020}{}}%
Glenza, J. (2020). Covid cases increase across {US} as upper midwest sees rapid rise. \emph{The Guardian}. \url{https://www.theguardian.com/world/2020/oct/22/covid-cases-coronavirus-us-midwest}

\leavevmode\vadjust pre{\hypertarget{ref-goldstein_during_2022}{}}%
Goldstein, J. (2022). During the {Omicron} surge, {Black} {New} {Yorkers} were hospitalized at a rate more than twice that of white residents. \emph{The New York Times}. \url{https://www.nytimes.com/2022/03/03/nyregion/omicron-hospitalizations-new-york-city.html}

\leavevmode\vadjust pre{\hypertarget{ref-gonzalez_conservatism_2021}{}}%
Gonzalez, K. E., James, R., Bjorklund, E. T., \& Hill, T. D. (2021). Conservatism and infrequent mask usage: {A} study of {US} counties during the novel coronavirus ({COVID}-19) pandemic. \emph{Social Science Quarterly}, \emph{102}(5), 2368--2382. \url{https://doi.org/10.1111/ssqu.13025}

\leavevmode\vadjust pre{\hypertarget{ref-greenland_ecologic_2001}{}}%
Greenland, S. (2001). Ecologic versus individual-level sources of bias in ecologic estimates of contextual health effects. \emph{International Journal of Epidemiology}, \emph{30}(6), 1343--1350. \url{https://doi.org/10.1093/ije/30.6.1343}

\leavevmode\vadjust pre{\hypertarget{ref-grossman_political_2020}{}}%
Grossman, G., Kim, S., Rexer, J. M., \& Thirumurthy, H. (2020). Political partisanship influences behavioral responses to governors' recommendations for {COVID}-19 prevention in the {United} {States}. \emph{Proceedings of the National Academy of Sciences}, \emph{117}(39), 24144--24153. \url{https://doi.org/10.1073/pnas.2007835117}

\leavevmode\vadjust pre{\hypertarget{ref-hanage_covid-19_2020}{}}%
Hanage, W. P., Testa, C., Chen, J. T., Davis, L., Pechter, E., Seminario, P., Santillana, M., \& Krieger, N. (2020). {COVID}-19: {US} federal accountability for entry, spread, and inequities---lessons for the future. \emph{European Journal of Epidemiology}, \emph{35}(11), 995--1006. \url{https://doi.org/10.1007/s10654-020-00689-2}

\leavevmode\vadjust pre{\hypertarget{ref-health_resources__services_administration_defining_2022}{}}%
Health Resources \& Services Administration. (2022). \emph{Defining {Rural} {Population}}. \url{https://www.hrsa.gov/rural-health/about-us/what-is-rural}

\leavevmode\vadjust pre{\hypertarget{ref-hernandez_disparities_2022}{}}%
Hernandez, I., Dickson, S., Tang, S., Gabriel, N., Berenbrok, L. A., \& Guo, J. (2022). Disparities in distribution of {COVID}-19 vaccines across {US} counties: {A} geographic information system--based cross-sectional study. \emph{PLOS Medicine}, \emph{19}(7), e1004069. \url{https://doi.org/10.1371/journal.pmed.1004069}

\leavevmode\vadjust pre{\hypertarget{ref-hill_covid-19_2022}{}}%
Hill, L., \& Artiga, S. (2022). {COVID}-19 {Cases} and {Deaths} by {Race}/{Ethnicity}: {Current} {Data} and {Changes} {Over} {Time}. In \emph{KFF}. \url{https://www.kff.org/coronavirus-covid-19/issue-brief/covid-19-cases-and-deaths-by-race-ethnicity-current-data-and-changes-over-time/}

\leavevmode\vadjust pre{\hypertarget{ref-johnson_whites_2022}{}}%
Johnson, A., \& Keating, D. (2022). Whites now more likely to die from covid than {Blacks}: {Why} the pandemic shifted. In \emph{Washington Post}. \url{https://www.washingtonpost.com/health/2022/10/19/covid-deaths-us-race/}

\leavevmode\vadjust pre{\hypertarget{ref-kaashoek_evolving_2021}{}}%
Kaashoek, J., Testa, C., Chen, J., Stolerman, L., Krieger, N., Hanage, W. P., \& Santillana, M. (2021). \emph{The {Evolving} {Roles} of {US} {Political} {Partisanship} and {Social} {Vulnerability} in the {COVID}-19 {Pandemic} from {February} 2020 - {February} 2021} {[}\{SSRN\} \{Scholarly\} \{Paper\}{]}. \url{https://doi.org/10.2139/ssrn.3933453}

\leavevmode\vadjust pre{\hypertarget{ref-kim_covid-19_2020}{}}%
Kim, E. J., Marrast, L., \& Conigliaro, J. (2020). {COVID}-19: {Magnifying} the {Effect} of {Health} {Disparities}. \emph{Journal of General Internal Medicine}, \emph{35}(8), 2441--2442. \url{https://doi.org/10.1007/s11606-020-05881-4}

\leavevmode\vadjust pre{\hypertarget{ref-krieger_relationship_2022}{}}%
Krieger, N., Testa, C., Chen, J. T., Hanage, W. P., \& McGregor, A. J. (2022). Relationship of political ideology of {US} federal and state elected officials and key {COVID} pandemic outcomes following vaccine rollout to adults: {April} 2021--{March} 2022. \emph{The Lancet Regional Health -- Americas}, \emph{16}. \url{https://doi.org/10.1016/j.lana.2022.100384}

\leavevmode\vadjust pre{\hypertarget{ref-krieger_public_2016}{}}%
Krieger, N., Waterman, P. D., Spasojevic, J., Li, W., Maduro, G., \& Van Wye, G. (2016). Public {Health} {Monitoring} of {Privilege} and {Deprivation} {With} the {Index} of {Concentration} at the {Extremes}. \emph{American Journal of Public Health}, \emph{106}(2), 256--263. \url{https://doi.org/10.2105/AJPH.2015.302955}

\leavevmode\vadjust pre{\hypertarget{ref-lambert_most_2022}{}}%
Lambert, J. (2022). Most {Americans} {Don}'t {Know} {About} the {Omicron} {Covid} {Boosters}. In \emph{Grid News}. \url{https://www.grid.news/story/science/2022/10/04/most-americans-dont-know-about-the-omicron-covid-boosters-that-spells-trouble-for-the-coming-winter/}

\leavevmode\vadjust pre{\hypertarget{ref-leonhardt_red_2021}{}}%
Leonhardt, D. (2021). Red {Covid}. \emph{The New York Times}. \url{https://www.nytimes.com/2021/09/27/briefing/covid-red-states-vaccinations.html}

\leavevmode\vadjust pre{\hypertarget{ref-mayo_clinic_history_2022}{}}%
Mayo Clinic. (2022). History of {COVID}-19: {Outbreaks} and {Vaccine} {Timeline}. In \emph{MayoClinic.org}. \url{https://www.mayoclinic.org/coronavirus-covid-19/history-disease-outbreaks-vaccine-timeline/covid-19}

\leavevmode\vadjust pre{\hypertarget{ref-mit_election_data_and_science_lab_county_2022}{}}%
MIT Election Data and Science Lab. (2022). \emph{County {Presidential} {Election} {Returns} 2000-2020}. Harvard Dataverse. \url{https://doi.org/10.7910/DVN/VOQCHQ}

\leavevmode\vadjust pre{\hypertarget{ref-mollalo_gis-based_2020}{}}%
Mollalo, A., Vahedi, B., \& Rivera, K. M. (2020). {GIS}-based spatial modeling of {COVID}-19 incidence rate in the continental {United} {States}. \emph{Science of The Total Environment}, \emph{728}, 138884. \url{https://doi.org/10.1016/j.scitotenv.2020.138884}

\leavevmode\vadjust pre{\hypertarget{ref-moreland_timing_2020}{}}%
Moreland, A., Herlihy, C., Tynan, M. A., Sunshine, G., McCord, R. F., Hilton, C., Poovey, J., Werner, A. K., Jones, C. D., Fulmer, E. B., Gundlapalli, A. V., Strosnider, H., Potvien, A., García, M. C., Honeycutt, S., Baldwin, G., CDC Public Health Law Program, CDC COVID-19 Response Team, \& Mitigation Policy Analysis Unit. (2020). Timing of {State} and {Territorial} {COVID}-19 {Stay}-at-{Home} {Orders} and {Changes} in {Population} {Movement} --- {United} {States}, {March} 1--{May} 31, 2020. \emph{MMWR. Morbidity and Mortality Weekly Report}, \emph{69}. \url{https://doi.org/10.15585/mmwr.mm6935a2}

\leavevmode\vadjust pre{\hypertarget{ref-murthy_disparities_2021}{}}%
Murthy, B. P. (2021). Disparities in {COVID}-19 {Vaccination} {Coverage} {Between} {Urban} and {Rural} {Counties} --- {United} {States}, {December} 14, 2020--{April} 10, 2021. \emph{MMWR. Morbidity and Mortality Weekly Report}, \emph{70}. \url{https://doi.org/10.15585/mmwr.mm7020e3}

\leavevmode\vadjust pre{\hypertarget{ref-national_bureau_of_economic_research_measuring_2021}{}}%
National Bureau of Economic Research. (2021). Measuring the {Virus} {Risk} of {Essential} {Workers} and {Dependents}. In \emph{NBER}. \url{https://www.nber.org/digest/202103/measuring-virus-risk-essential-workers-and-dependents}

\leavevmode\vadjust pre{\hypertarget{ref-national_conference_of_state_legislatures_covid-19_2021}{}}%
National Conference of State Legislatures. (2021). {COVID}-19: {Essential} {Workers} in the {States}. In \emph{COVID-19: Essential Workers in the States}. \url{https://www.ncsl.org/research/labor-and-employment/covid-19-essential-workers-in-the-states.aspx}

\leavevmode\vadjust pre{\hypertarget{ref-office_of_the_commissioner_coronavirus_2022}{}}%
Office of the Commissioner. (2022). Coronavirus ({COVID}-19) {Update}: {FDA} {Authorizes} {Moderna}, {Pfizer}-{BioNTech} {Bivalent} {COVID}-19 {Vaccines} for {Use} as a {Booster} {Dose}. In \emph{FDA}. \url{https://www.fda.gov/news-events/press-announcements/coronavirus-covid-19-update-fda-authorizes-moderna-pfizer-biontech-bivalent-covid-19-vaccines-use}

\leavevmode\vadjust pre{\hypertarget{ref-okonkwo_covid-19_2021}{}}%
Okonkwo, N. E., Aguwa, U. T., Jang, M., Barré, I. A., Page, K. R., Sullivan, P. S., Beyrer, C., \& Baral, S. (2021). {COVID}-19 and the {US} response: Accelerating health inequities. \emph{BMJ Evidence-Based Medicine}, \emph{26}(4), 176--179. \url{https://doi.org/10.1136/bmjebm-2020-111426}

\leavevmode\vadjust pre{\hypertarget{ref-ortega_ending_2021}{}}%
Ortega, A. N., \& Roby, D. H. (2021). Ending {Structural} {Racism} in the {US} {Health} {Care} {System} to {Eliminate} {Health} {Care} {Inequities}. \emph{JAMA}, \emph{326}(7), 613--615. \url{https://doi.org/10.1001/jama.2021.11160}

\leavevmode\vadjust pre{\hypertarget{ref-park_covid-19_2021}{}}%
Park, Y. M., Kearney, G. D., Wall, B., Jones, K., Howard, R. J., \& Hylock, R. H. (2021). {COVID}-19 {Deaths} in the {United} {States}: {Shifts} in {Hot} {Spots} over the {Three} {Phases} of the {Pandemic} and the {Spatiotemporally} {Varying} {Impact} of {Pandemic} {Vulnerability}. \emph{International Journal of Environmental Research and Public Health}, \emph{18}(17), 8987. \url{https://doi.org/10.3390/ijerph18178987}

\leavevmode\vadjust pre{\hypertarget{ref-press_inequities_2021}{}}%
Press, V. G., Huisingh-Scheetz, M., \& Arora, V. M. (2021). Inequities in {Technology} {Contribute} to {Disparities} in {COVID}-19 {Vaccine} {Distribution}. \emph{JAMA Health Forum}, \emph{2}(3), e210264. \url{https://doi.org/10.1001/jamahealthforum.2021.0264}

\leavevmode\vadjust pre{\hypertarget{ref-rader_spatial_2021}{}}%
Rader, B., Astley, C. M., Sewalk, K., Delamater, P. L., Cordiano, K., Wronski, L., Rivera, J. M., Hallberg, K., Pera, M. F., Cantor, J., Whaley, C. M., Bravata, D. M., \& Brownstein, J. S. (2021). \emph{Spatial {Accessibility} {Modeling} of {Vaccine} {Deserts} as {Barriers} to {Controlling} {SARS}-{CoV}-2}. medRxiv. \url{https://doi.org/10.1101/2021.06.09.21252858}

\leavevmode\vadjust pre{\hypertarget{ref-rapp_statelevel_2022}{}}%
Rapp, K. S., Volpe, V. V., Hale, T. L., \& Quartararo, D. F. (2022). State--{Level} {Sexism} and {Gender} {Disparities} in {Health} {Care} {Access} and {Quality} in the {United} {States}. \emph{Journal of Health and Social Behavior}, \emph{63}(1), 2--18. \url{https://doi.org/10.1177/00221465211058153}

\leavevmode\vadjust pre{\hypertarget{ref-saha_neighborhood-level_2020}{}}%
Saha, S., \& Feldman, J. M. (2020). \emph{Neighborhood-level {Racial}/{Ethnic} and {Economic} {Inequities} in {COVID}-19 {Burden} {Within} {Urban} {Areas} in the {US} and {Canada}}. medRxiv. \url{https://doi.org/10.1101/2020.12.07.20241018}

\leavevmode\vadjust pre{\hypertarget{ref-scott_these_2020}{}}%
Scott, D. (2020). These 4 {Midwestern} states are seeing worrying {Covid}-19 spikes. In \emph{Vox}. \url{https://www.vox.com/2020/9/2/21418812/covid-19-coronavirus-us-cases-midwest-surge}

\leavevmode\vadjust pre{\hypertarget{ref-shumaker_covid-19_2020}{}}%
Shumaker, L., \& Wu, T. (2020). {COVID}-19 cases surge in {U}.{S}. {Midwest} and {Northeast}. \emph{Reuters}. \url{https://www.reuters.com/article/us-health-coronavirus-usa-trends-idUSKBN26Q30B}

\leavevmode\vadjust pre{\hypertarget{ref-sigler_socio-spatial_2021}{}}%
Sigler, T., Mahmuda, S., Kimpton, A., Loginova, J., Wohland, P., Charles-Edwards, E., \& Corcoran, J. (2021). The socio-spatial determinants of {COVID}-19 diffusion: The impact of globalisation, settlement characteristics and population. \emph{Globalization and Health}, \emph{17}(1), 56. \url{https://doi.org/10.1186/s12992-021-00707-2}

\leavevmode\vadjust pre{\hypertarget{ref-sonderlund_racialized_2022}{}}%
Sonderlund, A. L., Charifson, M., Schoenthaler, A., Carson, T., \& Williams, N. J. (2022). Racialized economic segregation and health outcomes: {A} systematic review of studies that use the {Index} of {Concentration} at the {Extremes} for race, income, and their interaction. \emph{PLOS ONE}, \emph{17}(1), e0262962. \url{https://doi.org/10.1371/journal.pone.0262962}

\leavevmode\vadjust pre{\hypertarget{ref-sugg_mapping_2021}{}}%
Sugg, M. M., Spaulding, T. J., Lane, S. J., Runkle, J. D., Harden, S. R., Hege, A., \& Iyer, L. S. (2021). Mapping community-level determinants of {COVID}-19 transmission in nursing homes: {A} multi-scale approach. \emph{Science of The Total Environment}, \emph{752}, 141946. \url{https://doi.org/10.1016/j.scitotenv.2020.141946}

\leavevmode\vadjust pre{\hypertarget{ref-testa_public_2022}{}}%
Testa, C., Chen, J., Hall, E., Javadi, D., Morgan, J., Rushovich, T., Saha, S., Waterman, P., \& Krieger, N. (2022). \emph{Public {Health} {Disparities} {Geocoding} {Project} 2.0 {Training} {Manual}}. \url{https://phdgp.github.io/PHDGP2.0/index.html}

\leavevmode\vadjust pre{\hypertarget{ref-the_lancet_plight_2020}{}}%
The Lancet. (2020). The plight of essential workers during the {COVID}-19 pandemic. \emph{The Lancet}, \emph{395}(10237), 1587. \url{https://doi.org/10.1016/S0140-6736(20)31200-9}

\leavevmode\vadjust pre{\hypertarget{ref-the_new_york_times_coronavirus_2021}{}}%
The New York Times. (2021). \emph{Coronavirus ({Covid}-19) {Data} in the {United} {States}}. The New York Times. \url{https://github.com/nytimes/covid-19-data}

\leavevmode\vadjust pre{\hypertarget{ref-the_white_house_executive_2021}{}}%
The White House. (2021). Executive {Order} {On} {Advancing} {Racial} {Equity} and {Support} for {Underserved} {Communities} {Through} the {Federal} {Government}. In \emph{The White House}. \url{https://www.whitehouse.gov/briefing-room/presidential-actions/2021/01/20/executive-order-advancing-racial-equity-and-support-for-underserved-communities-through-the-federal-government/}

\leavevmode\vadjust pre{\hypertarget{ref-the_white_house_advancing_2022}{}}%
The White House. (2022). Advancing {Equity} and {Racial} {Justice} {Through} the {Federal} {Government}. In \emph{The White House}. \url{https://www.whitehouse.gov/equity/}

\leavevmode\vadjust pre{\hypertarget{ref-thompson_covid-19_2020}{}}%
Thompson, C. N. (2020). {COVID}-19 {Outbreak} --- {New} {York} {City}, {February} 29--{June} 1, 2020. \emph{MMWR. Morbidity and Mortality Weekly Report}, \emph{69}. \url{https://doi.org/10.15585/mmwr.mm6946a2}

\leavevmode\vadjust pre{\hypertarget{ref-noauthor_united_2022}{}}%
United {States} {COVID} - {Coronavirus} {Statistics} - {Worldometer}. (2022). In \emph{Worldometers}. \url{https://www.worldometers.info/coronavirus/country/us/}

\leavevmode\vadjust pre{\hypertarget{ref-us_department_of_commerce_economics_and_statistics_administration_census_2000}{}}%
U.S. Department of Commerce Economics and Statistics Administration, \& U.S. Census Bureau. (2000). \emph{Census {Regions} and {Divisions} of the {United} {States}}. \url{https://www2.census.gov/geo/pdfs/maps-data/maps/reference/us_regdiv.pdf}

\leavevmode\vadjust pre{\hypertarget{ref-us_census_bureau_nation_2022}{}}%
US Census Bureau. (2022). Nation {Continues} to {Age} as {It} {Becomes} {More} {Diverse}. In \emph{Census.gov}. \url{https://www.census.gov/newsroom/press-releases/2022/population-estimates-characteristics.html}

\leavevmode\vadjust pre{\hypertarget{ref-us_census_bureau_2020_2021}{}}%
US Census Bureau. (2021). 2020 {Census}: {Redistricting} {File} ({Public} {Law} 94-171) {Dataset}. In \emph{Census.gov}. \url{https://www.census.gov/data/datasets/2020/dec/2020-census-redistricting-summary-file-dataset.html}

\leavevmode\vadjust pre{\hypertarget{ref-us_census_bureau_american_2020}{}}%
US Census Bureau. (2020). American {Community} {Survey} 2015-2019 5-{Year} {Data} {Release}. In \emph{Census.gov}. \url{https://www.census.gov/newsroom/press-kits/2020/acs-5-year.html}

\leavevmode\vadjust pre{\hypertarget{ref-walker_tidycensus_2022}{}}%
Walker, K., \& Herman, M. (2022). \emph{Tidycensus: {Load} {US} {Census} {Boundary} and {Attribute} {Data} as 'tidyverse' and 'sf'-{Ready} {Data} {Frames}}.

\leavevmode\vadjust pre{\hypertarget{ref-wei_risk_2022}{}}%
Wei, C.-F., Lan, F.-Y., Hsu, Y.-T., Lowery, N., Dibona, L., Akkeh, R., Kales, S. N., \& Yang, J. (2022). Risk of {SARS}-{CoV}-2 {Infection} {Among} {Essential} {Workers} in a {Community}-{Based} {Cohort} in the {United} {States}. \emph{Frontiers in Public Health}, \emph{10}. \url{https://www.frontiersin.org/articles/10.3389/fpubh.2022.878208}

\leavevmode\vadjust pre{\hypertarget{ref-whitehead_outness_2016}{}}%
Whitehead, J., Shaver, J., \& Stephenson, R. (2016). Outness, {Stigma}, and {Primary} {Health} {Care} {Utilization} among {Rural} {LGBT} {Populations}. \emph{PLOS ONE}, \emph{11}(1), e0146139. \url{https://doi.org/10.1371/journal.pone.0146139}

\leavevmode\vadjust pre{\hypertarget{ref-whittle_ecological_2020}{}}%
Whittle, R. S., \& Diaz-Artiles, A. (2020). An ecological study of socioeconomic predictors in detection of {COVID}-19 cases across neighborhoods in {New} {York} {City}. \emph{BMC Medicine}, \emph{18}(1), 271. \url{https://doi.org/10.1186/s12916-020-01731-6}

\leavevmode\vadjust pre{\hypertarget{ref-wood_mgcv_2022}{}}%
Wood, S. (2022). \emph{Mgcv: {Mixed} {GAM} {Computation} {Vehicle} with {Automatic} {Smoothness} {Estimation}}. \url{https://CRAN.R-project.org/package=mgcv}

\leavevmode\vadjust pre{\hypertarget{ref-wood_generalized_2017}{}}%
Wood, S. N. (2017). \emph{Generalized {Additive} {Models}: {An} {Introduction} with {R}} (2nd ed.). Chapman \& Hall/CRC Texts in Statistical Science. \url{https://www.routledge.com/Generalized-Additive-Models-An-Introduction-with-R-Second-Edition/Wood/p/book/9781498728331}

\leavevmode\vadjust pre{\hypertarget{ref-wood_soap_2008}{}}%
Wood, S. N., Bravington, M. V., \& Hedley, S. L. (2008). Soap film smoothing. \emph{Journal of the Royal Statistical Society: Series B (Statistical Methodology)}, \emph{70}(5), 931--955. \url{https://doi.org/10.1111/j.1467-9868.2008.00665.x}

\end{CSLReferences}

\end{document}